\documentclass[preprint]{JHEP3} % 10pt is ignored!

\usepackage{epsfig,multicol,mcite,amsmath}
\usepackage{marvosym}
\newcommand\fverb{\setbox\fverbbox=\hbox\bgroup\verb}
\newcommand\fverbdo{\egroup\medskip\noindent%
                        \fbox{\unhbox\fverbbox}\ }
\newcommand\fverbit{\egroup\item[\fbox{\unhbox\fverbbox}]}
\newbox\fverbbox

\title{On direct measurement of the ${\mathbf W}$ production charge asymmetry at the LHC }

\author{ K.~Lohwasser\\
   Denys Wilkinson Building, Keble Road, Oxford, OX1 3RH, United Kingdom\\
and Physikalisches Institut,
Hermann-Herder-Str. 3,
D-79104 Freiburg,
Germany\\ 
   E-mail: \email{kristin.lohwasser@physik.uni-freiburg.de}}

\author{J.~Ferrando\\
        Denys Wilkinson Building, Keble Road, Oxford, OX1 3RH, United Kingdom\\
        E-mail: \email{j.ferrando1@physics.ox.ac.uk}}

\author{\c{C}.~I\c{s}sever\\
        Denys Wilkinson Building, Keble Road, Oxford, OX1 3RH, United Kingdom\\
        E-mail: \email{c.issever1@physics.ox.ac.uk}}

\received{\today}               %%
\accepted{\today}               %% These are for published papers.

\abstract{The prospects for making a direct measurement of the
$W$ production charge asymmetry at the LHC are discussed. A modification to the method used at the Tevatron is proposed for measurements at the LHC. The expected
sensitivity for such a measurement to parton distribution functions 
is compared to that for a measurement of the lepton charge asymmetry.
The direct measurement approach is found to be less useful for placing constraints on parton distribution functions at the LHC than a measurement of the lepton charge asymmetry. 
}


\begin{document} 

%\maketitle  IS IGNORED %%%%%%%%%%%

\section{Introduction}
The measurement of the $W$ production charge asymmetry at 
hadron-hadron colliders provides  important information about parton
distribution functions (PDFs). The simplest method is to measure the
asymmetry in the pseudorapidity distribution of the charged leptons
(the lepton asymmetry) arising from leptonic decay of the $W^{\pm}$ bosons. Such measurements
have been performed at the Tevatron, for both $W\rightarrow e \nu$ and $W \rightarrow \mu \nu$
events, by both the CDF~\cite{Abe:1998rv,Acosta:2005ud} and D0~\cite{Abazov:2007pm, Abazov:2008qv} collaborations and the data have been included in global fits of parton
distributions~\cite{Martin:2009iq,Pumplin:2002vw,Ball:2010de}. The asymmetry is mainly 
sensitive to valence quark distributions~\cite{Berger:1988tu}, providing complementary information 
to that obtained from measurements of structure functions in deep inelastic scattering~\cite{Chekanov:2008aa,Chekanov:2009gm,Adloff:2003uh,Aaron:2009kv,:2009wt}. Predictions for the asymmetry exist up to next-to-next-to-leading order in the strong coupling~\cite{Catani:2010en}.

The CDF collaboration has recently published a direct measurement of the $W$ 
 production charge asymmetry~\cite{Aaltonen:2009ta}. 
This measurement used a technique first proposed by Bodek 
{\it et al.}~\cite{Bodek:2007cz}, in which the $W$ boson rapidity is inferred event-by-event from 
the lepton four-momentum and the missing transverse momentum. This
inference enables a measurement of the $W$ asymmetry to be made at values
of rapidity where the uncertainty on the prediction from theory is much larger
 than the uncertainty on  the lepton 
asymmetry for any part of the pseudorapidity range. 

The inclusion of the CDF $W$ asymmetry~\cite{Aaltonen:2009ta} in PDF fits has not been straightforward: the data set gives the second largest contribution per data point to the $\chi^2$ of the NNPDF fit~\cite{Ball:2010de}. Attempts by the MSTW group~\cite{Thorne:2010kj} to include it in their fits have revealed tension with the electron asymmetry measured by D0~\cite{Abazov:2007pm}. To investigate this the CDF collaboration have produced the electron asymmetry for the same dataset as used in the $W$ asymmetry measurement with two interesting features revealed. Firstly this electron asymmetry agrees well with the D0 electron asymmetry and secondly, while the $W$ asymmetry agrees very well with the predictions made using the CTEQ6.6~\cite{Nadolsky:2008zw} PDF set, the CDF electron asymmetry disagrees with predictions made using the same PDF set. This apparent inconsistency may indicate biases which have not been accounted for in the method.

In this paper the feasibility of measuring directly the $W$ 
production charge asymmetry at the LHC is investigated for the first time using simulated data. The
 expected  sensitivity to the PDFs for such a measurement is compared
to that for a measurement of the lepton asymmetry. 

The paper is structured as follows: the simulated data samples used for this study are summarised in section~{\ref{sec:mc}}; the procedure used to measure the $W$ asymmetry is described in section~\ref{sec:analysis}. Problems specific to the LHC environment are described and modifications to the reconstruction procedure proposed in sections~\ref{sec:weighting} and~\ref{sec:perfweighting}. The performance of the modified method is studied in greater detail and the expected statistical and systematic uncertainties on the $W$ asymmetry are compared to the expected statistical uncertainties on the lepton asymmetry for different luminosities in section~\ref{sec:PerformanceNewScheme}. Finally, the implications of this study for early measurements of $W$ production charge asymmetries at the LHC are discussed in section~\ref{sec:comparisons}.

\section{Simulation of $\mathbf{W}$ production}
\label{sec:mc}
The studies in this paper are based on samples generated using the Monte Carlo (MC) generators {\sc Pythia} 8.120~\cite{Sjostrand:2007gs} and Herwig++ 2.4.0~\cite{Bahr:2008pv,Bahr:2008tf}. The {\sc Pythia} sample is a simulation of the process $f\bar{f'} \rightarrow W$, where $f$ and $f'$ are fermions, at leading order (LO) in the strong coupling.
The Herwig++ sample is a simulation of the $W$ production process at next-to-leading order (NLO) in the strong coupling. As such, it represents the state-of-the-art in simulation of Drell-Yan vector boson production. The positive weight NLO matching scheme (POWHEG)~\cite{Nason:2004rx,Frixione:2007vw} was used in the generation of the Herwig++ sample. This approach consistently combines the NLO calculation and parton shower simulation and was implemented in Herwig++ by Hamilton {\em et al.}~\cite{Hamilton:2008pd}.

The PDFs used for the generation and the size of the samples used are specified in the appropriate sections of the paper. 
No detector simulation is applied. The following fiducial cuts are used in the event selection in order simulate the detector acceptance:
\begin{itemize}
\item a cut on the transverse momentum of the lepton, $P^l_T>25$ GeV,
\item a cut on the missing transverse momentum (the transverse momentum of the neutrino) in the event $\displaystyle{\not} E_T>25$ GeV,
\item a cut on the lepton pseudorapidity $|\eta^l|<2.4$.
\end{itemize}
\section{Analysis Technique}
\label{sec:analysis}

The procedure used to extract the $W$ production charge asymmetry as a 
function of the $W$ boson rapidity, $y_W$,
follows closely that used by Bodek {\it et al.}~\cite{Bodek:2007cz}. It can
be broken down into the following steps:

\begin{enumerate}
\item \textbf{Calculation of the $W$ rapidity solutions}: For each reconstructed event, the two possible $W$ rapidity values (referred to as ``solutions'')  are calculated from the missing transverse momentum in the event and the charged lepton momentum using a constraint on the $W$ boson mass.

\item \textbf{Weighting of the $W$ rapidity solutions}: Event-by-event, the solutions are filled into separate $y_{W^+}$ and $y_{W^-}$ histograms, using weights calculated from MC input as described in section \ref{sec:weighting}. If only one physical solution is found, the solution is filled with a weight of 1.
Step 1 (calculation of the solutions) and step 2 (weighting of the solutions) are hereafter referred to as \textbf{full kinematic $W$ reconstruction}. 

\item \textbf{Acceptance corrections}: The rapidities of the kinematically fully reconstructed $W$ bosons are corrected in each bin of 
$y_W$ for the detector acceptance and also for biases of the full kinematic $W$ reconstruction arising from the weighting procedure (as discussed in section \ref{sec:weighting}). The acceptance corrections are calculated bin-by-bin as the ratio between the generated $y_{W^{\pm}}$ distributions after  cuts  and full kinematic $W$ reconstruction and the MC generated $W$ distributions before cuts:

\begin{equation}\small
\mathrm{\small Acc}_{\mathrm{(bin\,i})} = \frac{\mathrm{MC \, events}\,\mathrm{\,after\, cuts\, and} \mathrm{\,with\,full\, kinematic\,} W \mathrm{\,reconstruction} _{\mathrm{(bin\,i})} }{\mathrm{MC\,events\,before\,cuts}_{\mathrm{(bin\,i})} }
\end{equation}

The acceptance correction is applied by
 multiplying $1/\mathrm{Acc}_{\mathrm{(bin\,i})}$ with the content of bin i of the $y_W$ distributions of the fully reconstructed $W$ bosons. Step 3 yields the \textbf{reconstructed $y_W$ distributions}, extrapolated to the whole phase space.

\item \textbf{Compare reconstructed and MC-input $y_W$ distributions}:  If the reconstructed $y_W$ distributions and the MC input $y_W$ distributions are consistent with each other then the procedure stops (go to step 7). If they disagree, then the procedure must  be iterated, with the MC input reweighted to describe the reconstructed distribution (go to step 5). 

\item \textbf{Reweight input MC to reproduce the reconstructed $y_W$ distributions}: The input MC $y_W$ distributions without any
 cuts  applied and the $y_W$ distributions reconstructed in step 1-3 are compared. A reweighting factor is extracted bin-by-bin:
\begin{equation}
r {\mathrm{(bin\,\,i})} = \frac{\mathrm{experimentally\,\, determined\,\,}y_{W \,\,{\mathrm{(bin\,\,i})}}}{\mathrm{Input\,\, MC\,} y^{tru
e}_W {\mathrm{(bin\,\,i})} }
\end{equation}

Each event in the input MC is re-weighted with the event weight $r$ ($y^{true}_W$ (MC input)). 

\item \textbf{Recalculation of weights}: The weights used in step 2 
and the acceptance corrections (step 3) are recalculated from the reweighted MC input and the steps 1-4 are repeated until convergence. This iteration reduces the dependence on the MC input. 

\item \textbf{Measurement of the $W$ asymmetry:} When the reconstructed and modified input MC $y_W$ distributions  are in agreement within their statistical uncertainty, the $W$ asymmetry is considered to be determined.

\end{enumerate}

\section{Full kinematic $W$ reconstruction: weighting of the $W$ rapidity solutions}\label{sec:weighting}

In the weighting procedure~\cite{Bodek:2007cz} (step 2 from section \ref{sec:analysis}) the twofold ambiguity for $p_z^{\nu}$ and thus for $y_W$ is resolved statistically with the help of MC predictions. The possible rapidity solutions $s_1$ and $s_2$ are weighted with their respective probabilities ($P_1$ and $P_2$) and the total probability is normalised to unity, $\mathcal{P}=P_1+P_2=1$. The probabilities of each solution  occuring are derived, based on:

\begin{enumerate}
\item \textbf{Expected cross sections $d\sigma/dy_W$}: The cross sections $d\sigma/dy_W$ have a distinct behaviour as a function of $y_W$. The two different rapidity solutions can be compared as to which is more probable. The expected cross sections $d\sigma/dy_W$ predicted by Monte Carlo are used as probability density functions to determine this. 

\item \textbf{Expected lepton decay angle and anti-quark/quark ratios}: The lepton decay angle in the $W$ rest frame, $\cos \theta^*$, follows a $(1 \pm \cos\theta^*)^2$ distribution, which can be used as another probability density function to weight the two rapidity solutions. The sign of the $(1 \pm \cos\theta^*)^2$ distribution depends on the helicity of the lepton and the helicity of the incoming parton with highest-$x$, hereafter referred to as ``higher-$x$ \mbox{(anti-)quarks}''. Therefore in the $\cos\theta^*$ weighting, the probability density function is built from the contributions of events with higher-$x$ quarks $(1 \mp \cos\theta^*)^2$ and of events with higher-$x$ anti-quarks $(1 \pm \cos\theta^*)^2$. The probability density function for the weighting is constructed as the sum of the two decay distributions. The contribution of the higher-$x$ anti-quark events is adjusted using the parameter $\frac{\bar{q}}{q}$, which is defined as

\begin{equation}
\frac{\bar{q}}{q} (p_T^{W}, y_{W}) = \frac{\operatorname{number\ of}\,\operatorname{higher-}x\,\, \mathrm{\operatorname{anti-quarks}}}{\operatorname{number\ of}\,\operatorname{higher-}x\,\, \mathrm{quarks}}  (p_T^{W}, y_{W}) 
\end{equation}

The ratio $\frac{\bar{q}}{q}$ is a function of $p_T^{W}$ and $y_{W}$, it is determined from Monte Carlo simulations and therefore also depends on PDFs. Using $\frac{\bar{q}}{q}$, one can build the total probability density function from the sum of the two possible angular distributions for the LHC
\begin{eqnarray}
P(\cos\theta^*) &=& (1 \mp \cos\theta^*)^2  + \frac{\bar{q}}{q} (p_T^{W}, y_{W}) (1 \pm \cos\theta^*)^2 \label{eq:cosweight_lhc}
\end{eqnarray}
and similarly for the Tevatron~\cite{Bodek:2007cz}.
\end{enumerate}

The probability density function for the cross section and the decay angle can be combined to calculate weighting factors, $w_1$ and $w_2$, for $s_1$ and $s_2$, depending on the charge of the $W^{\pm}$:

\begin{equation}\label{eq:rapweighting}
w_{1,2} (W^{\pm}) = \frac{ P(\cos\theta^*_{l^{\pm},W^{\pm}}) \times (d\sigma/dy_{W^{\pm}_{1,2}})}{ \sum_{i=1}^{2}{\left( P(\cos\theta^*_{l^{\pm},W^{\pm}}) \times (d\sigma/dy_{W^{\pm}_i})     \right)   }  }.
\end{equation}

In a reconstructed event each of the solutions 1 and 2 is weighted by $w_{1}$ and $w_2$ respectively and filled with that weight into histograms of  $y_{W^{\pm}}$. 

\section{Performance of the kinematic $W$ reconstruction}\label{sec:perfweighting}
The performance of the weighting and reconstruction of the $W$, steps 1 and 2 in the iterative procedure, was studied. This study was 
performed using  8$\times10^6$ {\sc Pythia} MC events generated with the MSTW08  NLO PDF set~\cite{Martin:2009iq} for Tevatron ($p\bar{p}$, $\sqrt{s}=1.96$ TeV) and LHC  conditions ($pp$, $\sqrt{s}=14$ TeV). The analysis technique was tested using the same MC sample as MC input and as 'pseudo-data'. The transverse components of the neutrino's momentum, $p_x^{\nu}$ and $p_y^{\nu}$ were taken directly from the MC. No acceptance corrections were applied for this part of the study.

\subsection{Performance at the Tevatron}
\label{subsec:tevperf}
\begin{figure}
%%%%%%%%%%%%%%%%%%%%%%%%%%%%
\begin{center}
\includegraphics[width=0.49\textwidth]{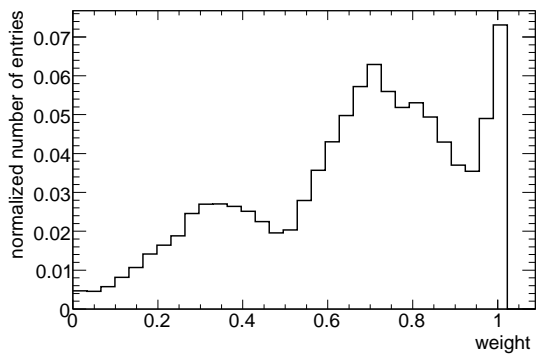}\\
\end{center}
\caption[Weight of solution closest to true $W^-$ rapidity at the Tevatron]{Normalised distributions of weights given to the rapidity solution that is closest to true $W^-$ rapidity in the Tevatron environment.}
\label{fig:fermilab_rightweight}
\end{figure}

Figure \ref{fig:fermilab_rightweight} shows the weight assigned to the rapidity solution  calculated in step 1 that was closest to the true rapidity. The figure shows the normalised distribution only for $W^-$ events but is very similar for $W^+$ events. The distribution exhibits a peak at 1 and lies mostly above 0.5, indicating that in the weighting procedure  the solution closest to the true $y_W$ value is given the larger weight. 

%xxxxxxxxxxxxx CALCULATED DISTRIBUTION

\begin{figure}
%%%%%%%%%%%%%%%%%%%%%%%%%%%%
\begin{minipage}[c]{0.49\textwidth}
\begin{center}
\includegraphics[width=0.99\textwidth]{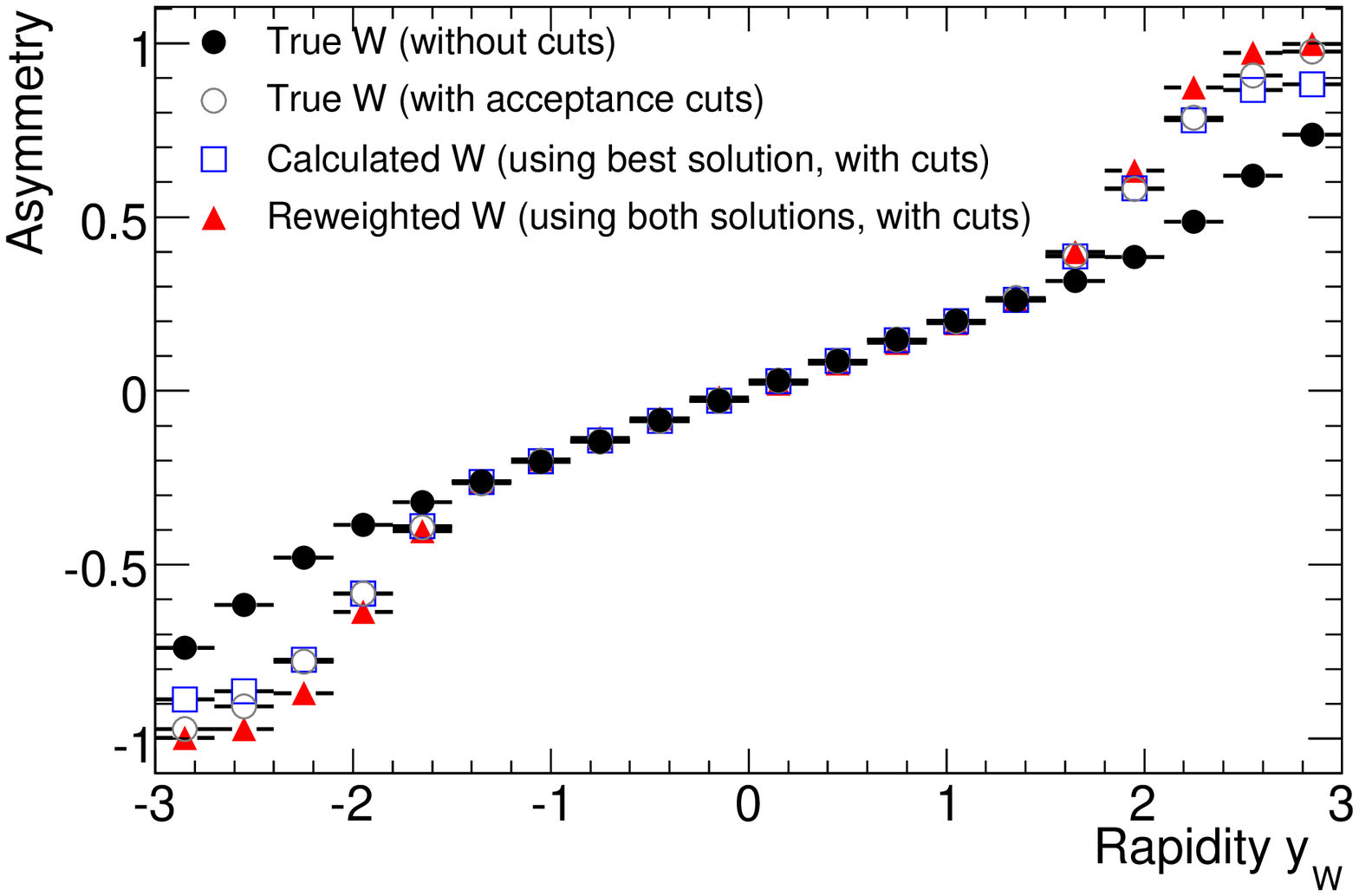}\\
\footnotesize {a) $W$ asymmetry  \linebreak
  }\\
\end{center}
\end{minipage}
%%%%%%%%%%%%%%%%%%%%%%%%%%%%
\begin{minipage}[c]{0.49\textwidth}
\begin{center}
\includegraphics[width=0.99\textwidth]{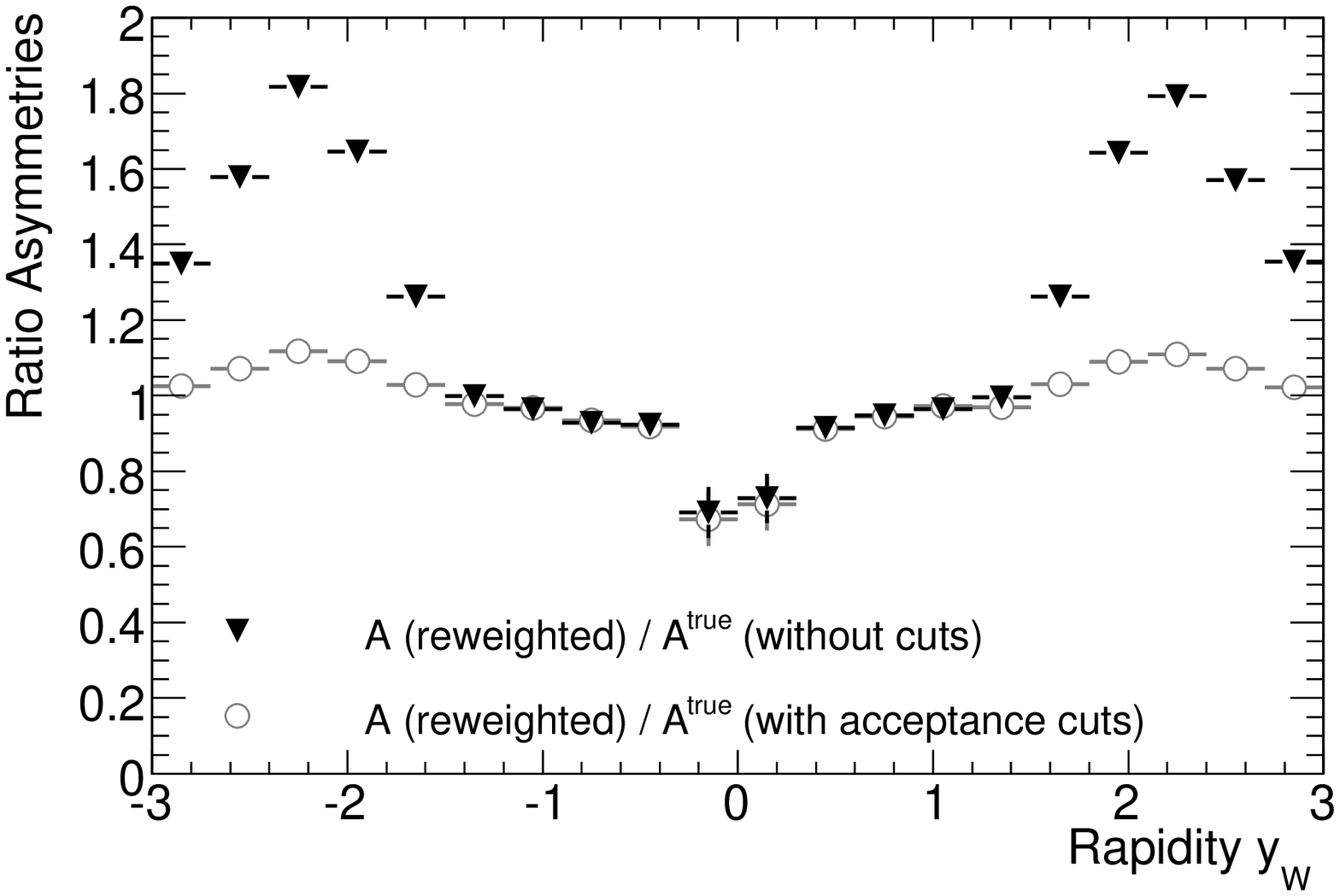}\\
\footnotesize {b) Ratio of reconstructed asymmetry and true asymmetries (with and without cuts)}
\end{center}
\end{minipage}
\caption[Asymmetry at the Tevatron]{This figure shows the asymmetry as evaluated at various stages of the iterative weighting procedure (a). In b) the ratios of reconstructed asymmetry and true asymmetries (before and after applying cuts) are depicted. }
\label{fig:fermilab_weighting2}
\end{figure}

Figure \ref{fig:fermilab_weighting2} a) shows the asymmetry distributions for 
the Tevatron. Shown are the true $W$ asymmetry distributions from the MC with and without cuts. Also depicted is the $W$ asymmetry that arises when only the $W$ solution closest to the true $y_W$ value is used and the $W$ asymmetry based on the full kinematic reconstruction. All asymmetries are in agreement up to $y_W=1.5$, beyond that there is  some disagreement between the reconstructed and the true $W$ asymmetry. This  can be  better seen in figure \ref{fig:fermilab_weighting2} b), which shows  the ratio of the fully reconstructed $W$ asymmetry (without acceptance corrections) and the true asymmetry. The  difference between the reconstructed and the true asymmetry in the region defined by the acceptance cuts  is only around 20\%. The  true asymmetry before cuts is at most 80\% to 100\% larger than  the reconstructed asymmetry. Hence it can be seen that acceptance corrections  correct not only for genuine detector acceptance effects but also for biases of the full kinematic reconstruction. However, the former are larger and more important than the latter.

\subsection{Performance at the LHC}

It is not immediately obvious that it is possible to transfer the full kinematic  reconstruction technique to the LHC because of crucial differences between 
the Tevatron and LHC environments. The Tevatron is a $p\bar{p}$ collider 
operating at centre-of-mass energy $\sqrt{s}=$1.96 TeV, while the LHC is due to collide $pp$ beams 
at $\sqrt{s}$ up to 14 TeV. These differences pose a major challenge when applying 
the weighting method to the LHC environment. Colliding $pp$ beams, will break
 the mirror symmetry between $W^+$ and $W^-$. The asymmetry in $pp$ collisions is the same at positive and negative rapidities.  At the LHC  $W^+$  bosons are 
produced almost twice as often as $W^-$ and generally with a larger boost along the $Z$ axis. Hence the performance of the method will differ in the two environments.

The full kinematic reconstruction and the weights used were studied using a {\sc Pythia} MC with the MSTW08 NLO PDF set. As already described above, the two solutions calculated in step 1, were weighted in step 2 using two basic considerations.

\begin{figure}
%%%%%% %%%%%%%%%%%%%%%%%%%%%%edge effect 
\begin{minipage}[c]{0.5\textwidth}
\begin{center}
\includegraphics[width=0.9\textwidth]{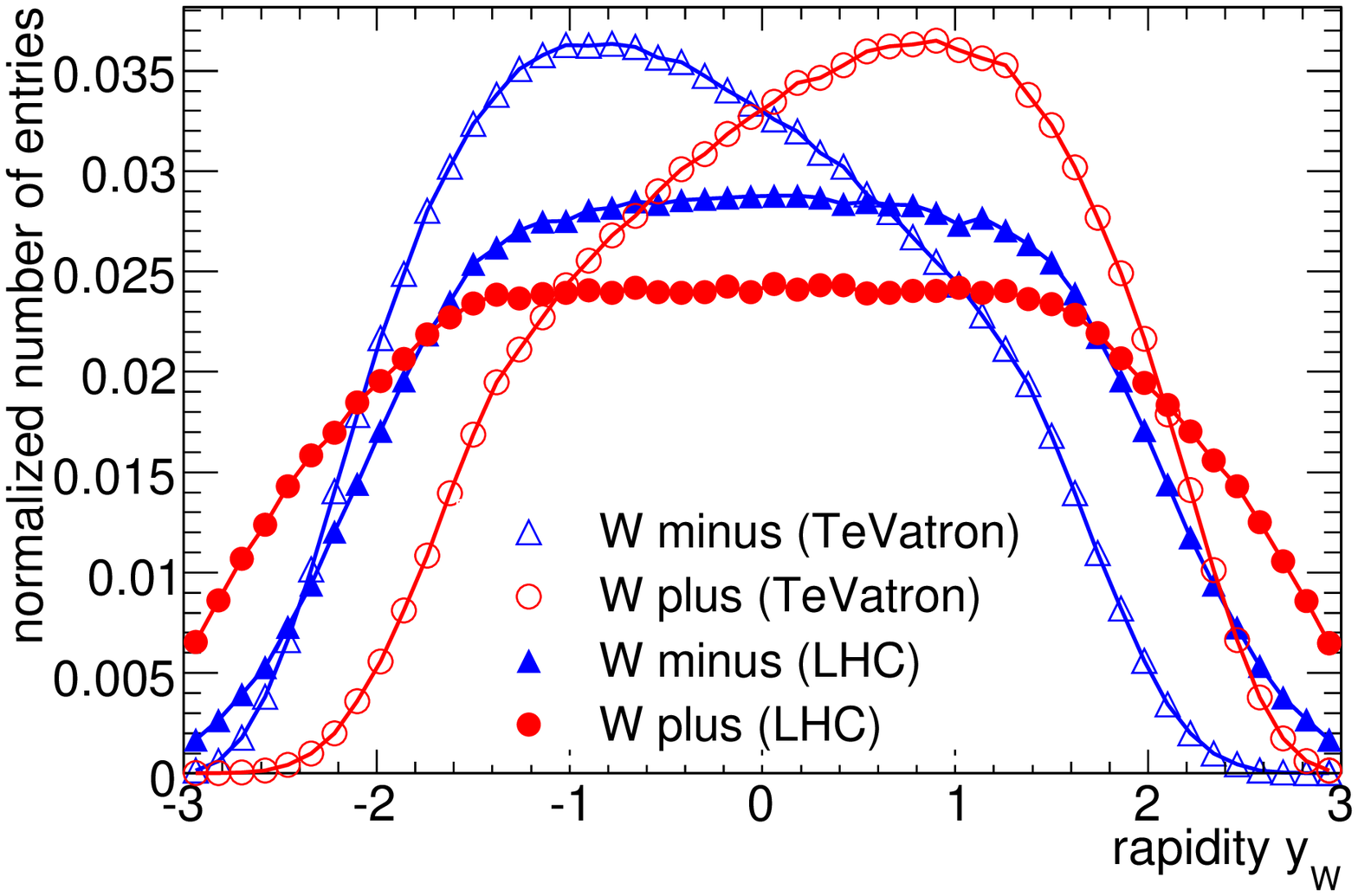}\\
\footnotesize {a) Normalised $y_{W^\pm}$ distributions }
\end{center}
\end{minipage}
%%%%%%%%%%%%%%%%%%%%%%%%%%%%
\parbox{0.49\textwidth}{
\center
\begin{minipage}{0.4\textwidth}

\caption[Problems of the weighting procedure at the LHC]{ {Problems of the weighting procedure at the LHC: a) The normalised $y_{W^\pm}$ distributions are flatter at the LHC (full markers) compared to the Tevatron (open markers). The ratios $\bar{q}/q$ used in the weighting procedure at the LHC are shown for $W^-$ in b) and for $W^+$ in c). For both charges, the ratio is considerably larger than 0.25, which is the maximal value at the Tevatron~\cite{Bodek:2007cz}.  \label{fig:lhc_qqratios} }}
\end{minipage}
}
%%%%%%%%%%%%%%%%%%%%%%%%%%%%
\begin{minipage}[c]{0.5\textwidth}
\begin{center}
\includegraphics[width=0.9\textwidth]{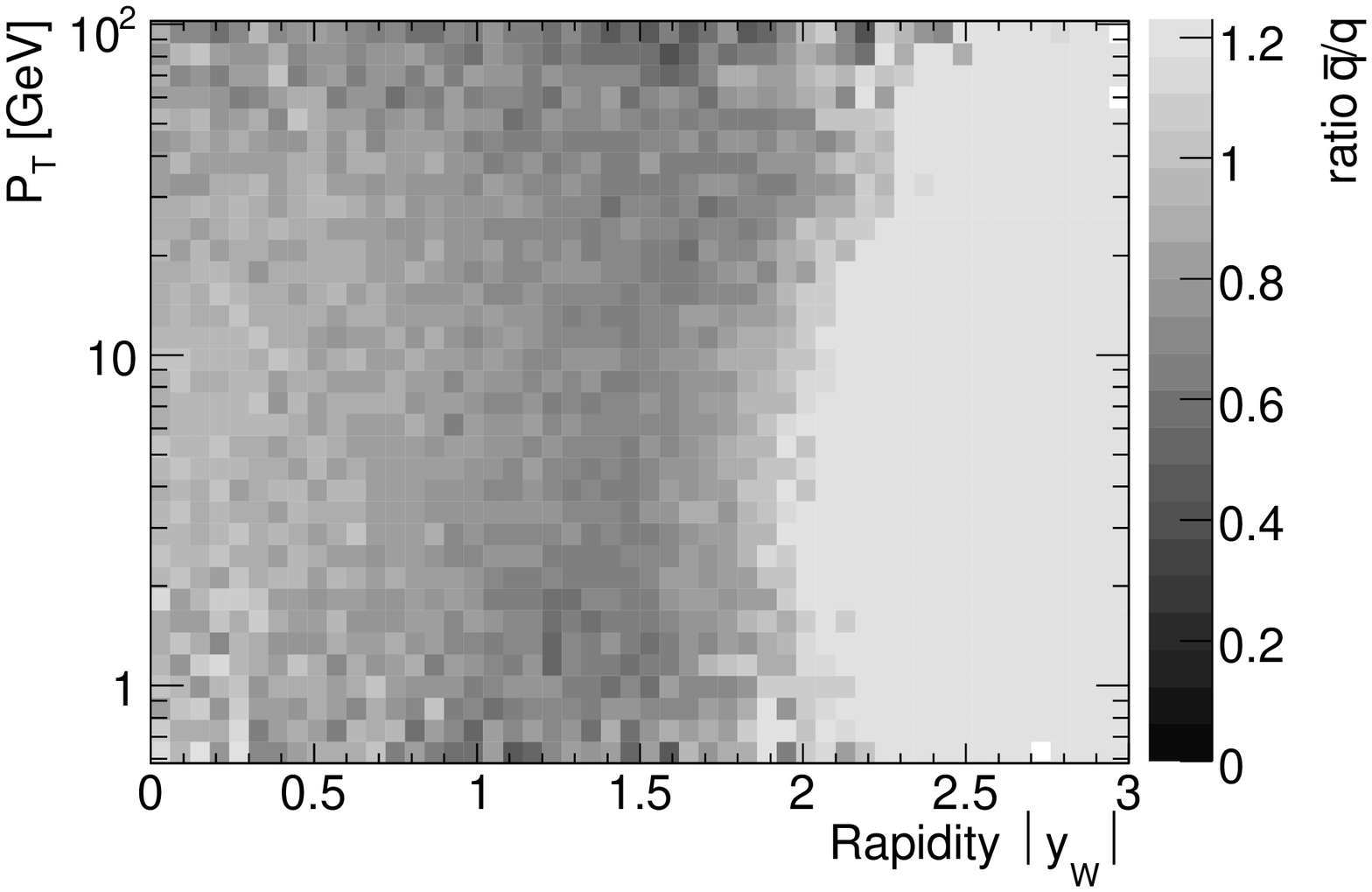}\\
\footnotesize {b) $\bar{q}/q$ ratio for $W^-$}
\end{center}
\end{minipage}
%%%%%%%%%%%%%%%%%%%%%%%%%%%%
\begin{minipage}[c]{0.5\textwidth}
\begin{center}
\includegraphics[width=0.9\textwidth]{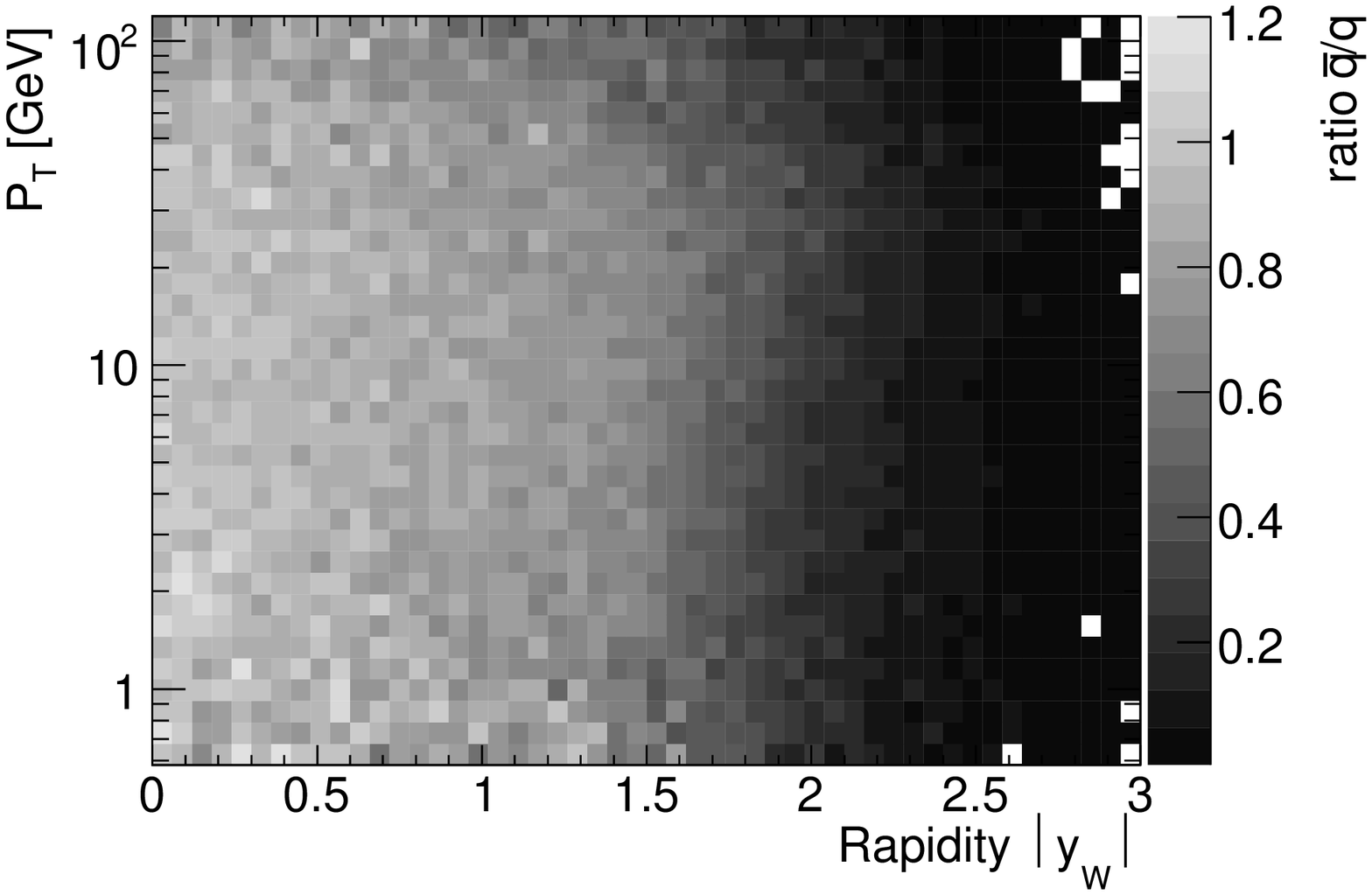}\\
\footnotesize {c) $\bar{q}/q$ ratio for $W^+$}
\end{center}
\end{minipage}
\end{figure}

\begin{figure}
\begin{minipage}[c]{0.5\textwidth}
\begin{center}
\includegraphics[width=0.9\textwidth]{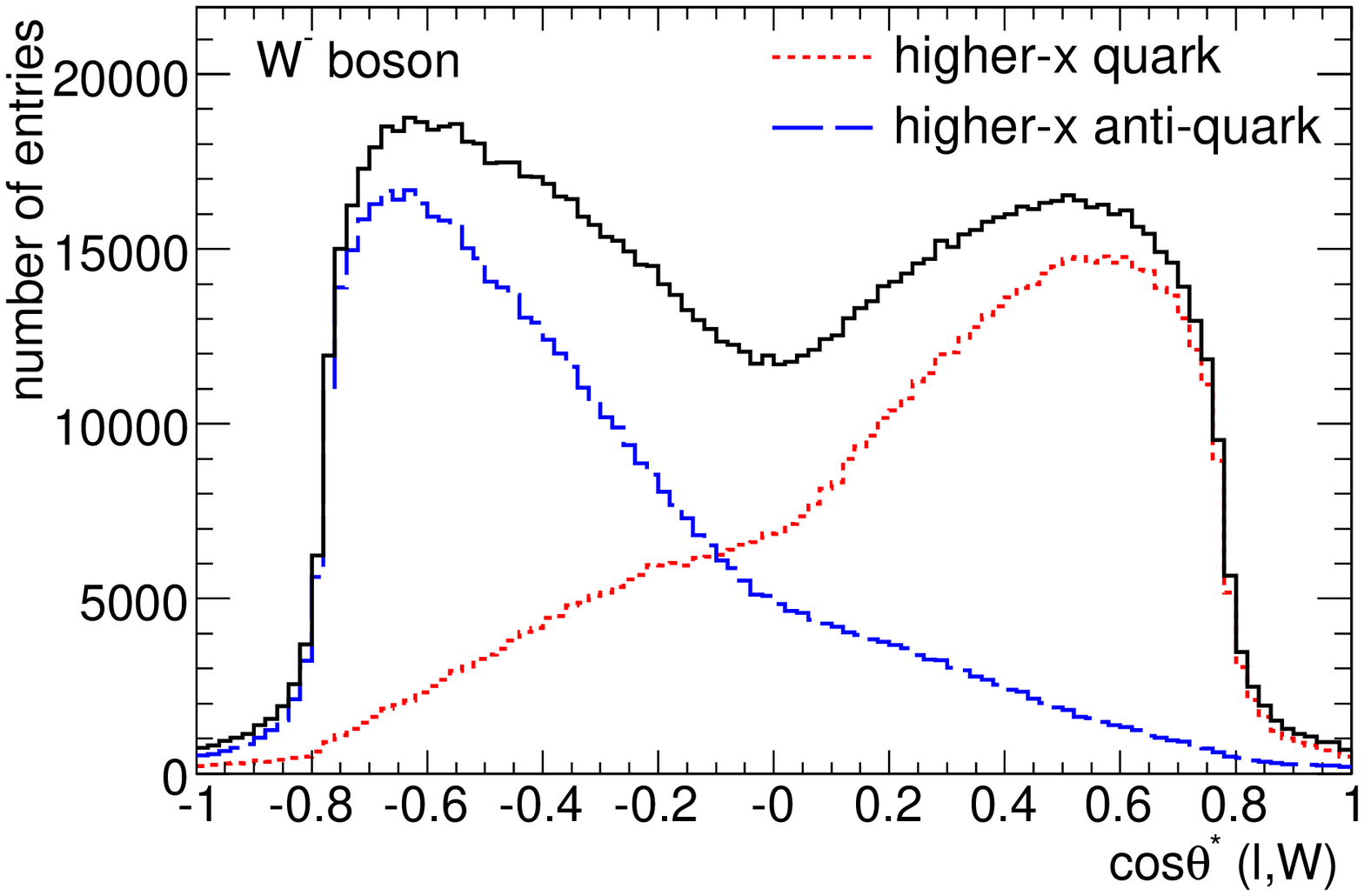}\\
\footnotesize {a) $\cos\theta^*$ distribution for $W^-$}
\end{center}
\end{minipage}
%%%%%%%%%%%%%%%%%%%%%%%%%%%%
\begin{minipage}[c]{0.5\textwidth}
\begin{center}
\includegraphics[width=0.9\textwidth]{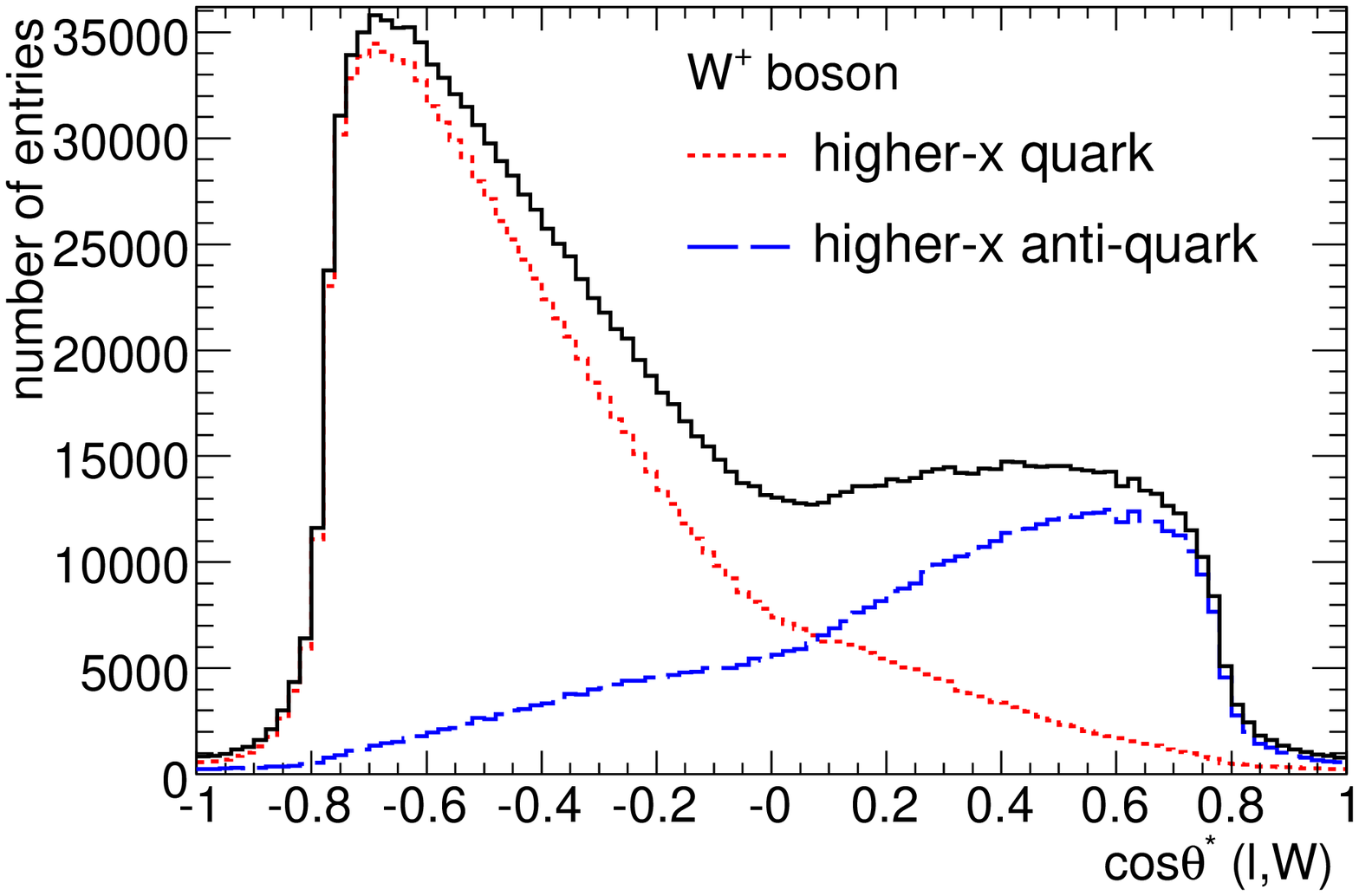}\\
\footnotesize {b) $\cos\theta^*$ distribution for $W^+$}
\end{center}
\end{minipage}
\caption[$\cos\theta^*$ distributions at the LHC]{$\cos\theta^*$ distributions at the LHC. 
The total of the higher-$x$ quark (dotted line) and higher-$x$ anti-quark (dashed line) distributions is shown. 
}

\label{fig:lhc_cosweights}
\end{figure}

\begin{itemize}
\item \textbf{Shape of cross section as function of $y_W$:} At the LHC the cross section for $W$ production is much flatter and expands over a much larger range of $y_W$ than at the Tevatron. This is shown in figure \ref{fig:lhc_qqratios}. There is no peak in the $y_W$ distributions for central rapidities and the cross section begins to fall only above $y_W>1.5-2.0$. Therefore, if both reconstructed solutions are within $-1.5<y_W<1.5$, both are equally probable. In particular, the reconstruction of $W^+$ rapidities suffers from this fact, since the longitudinal boost of $W^+$ bosons is generally larger than the boost of $W^-$, and therefore the rapidity distribution extends to higher $y_W$. 

\item \textbf{Ratios of leading $\bar{q}$ versus leading $q$:} A further problem is the contribution of $W$s produced with a higher-$x$ anti-quark. If the higher-$x$ parton participating in the Drell-Yan $W$ production is an anti-quark, the $W$ is no longer produced with a boost parallel to the incoming quark, but antiparallel to the incoming quark. This introduces a change of sign in the expected $\cos\theta^*$ decay angle. In the weighting procedure this is accounted for by constructing the $\cos\theta^*$ weight such that it combines a weight for leading quark and leading anti-quark $W$ production according to their relative contributions by using the ratio $\bar{q}/q$ as a function of $p_T^W$ and $y_W$, as in equation \ref{eq:rapweighting}. At Tevatron centre of mass energies, leading quark production is indeed the most probable process, with $\bar{q}/q$=0.25 at most. At the LHC however, the ratio $\bar{q}/q$  even extends to values as high as 1 as shown in figure \ref{fig:lhc_qqratios}. This introduces a
 strong ambiguity to the $\cos\theta^*$ weighting. This effect is
 particularly problematic for $W^-$ production because the ratio of anti-quark induced processes is higher than for $W^+$ production. The effect is also shown in figure \ref{fig:lhc_cosweights}, where the $\cos\theta^*$ distributions of $W^-$ (left) and $W^+$ (right) bosons are shown for $W$ production at the LHC. 
For the $W^-$ distributions, the contribution of 
higher-$x$ anti-quarks is especially large and two solutions with
 $\cos\theta^{*,1}\sim-\cos\theta^{*,2}$ cannot be distinguished.
 Solutions with $\cos\theta^{*,2}\sim-\cos\theta^{*,1}$ or 
$\cos\theta^{*,2}\sim\cos\theta^{*,1}$ are in fact most common and in these cases both solutions will be weighted with factors of about 0.5.
\end{itemize}

\begin{figure}
\begin{minipage}[c]{0.5\textwidth}
\begin{center}
\includegraphics[width=0.9\textwidth]{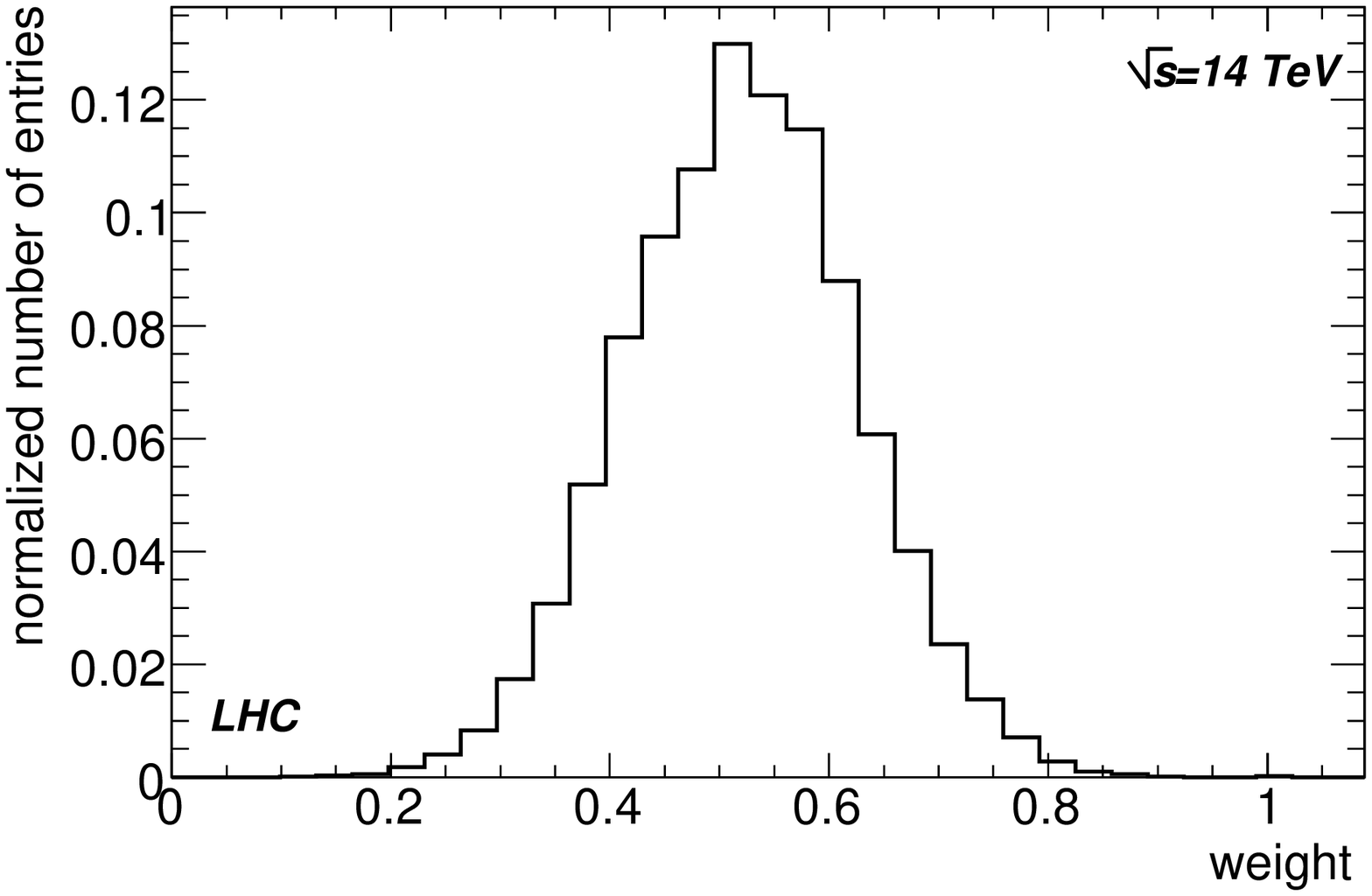}\\
\footnotesize {a) Weight given to the right solution for $W^-$}
\end{center}
\end{minipage}
%%%%%%%%%%%%%%%%%%%%%%%%%%%%
\begin{minipage}[c]{0.5\textwidth}
\begin{center}
\includegraphics[width=0.9\textwidth]{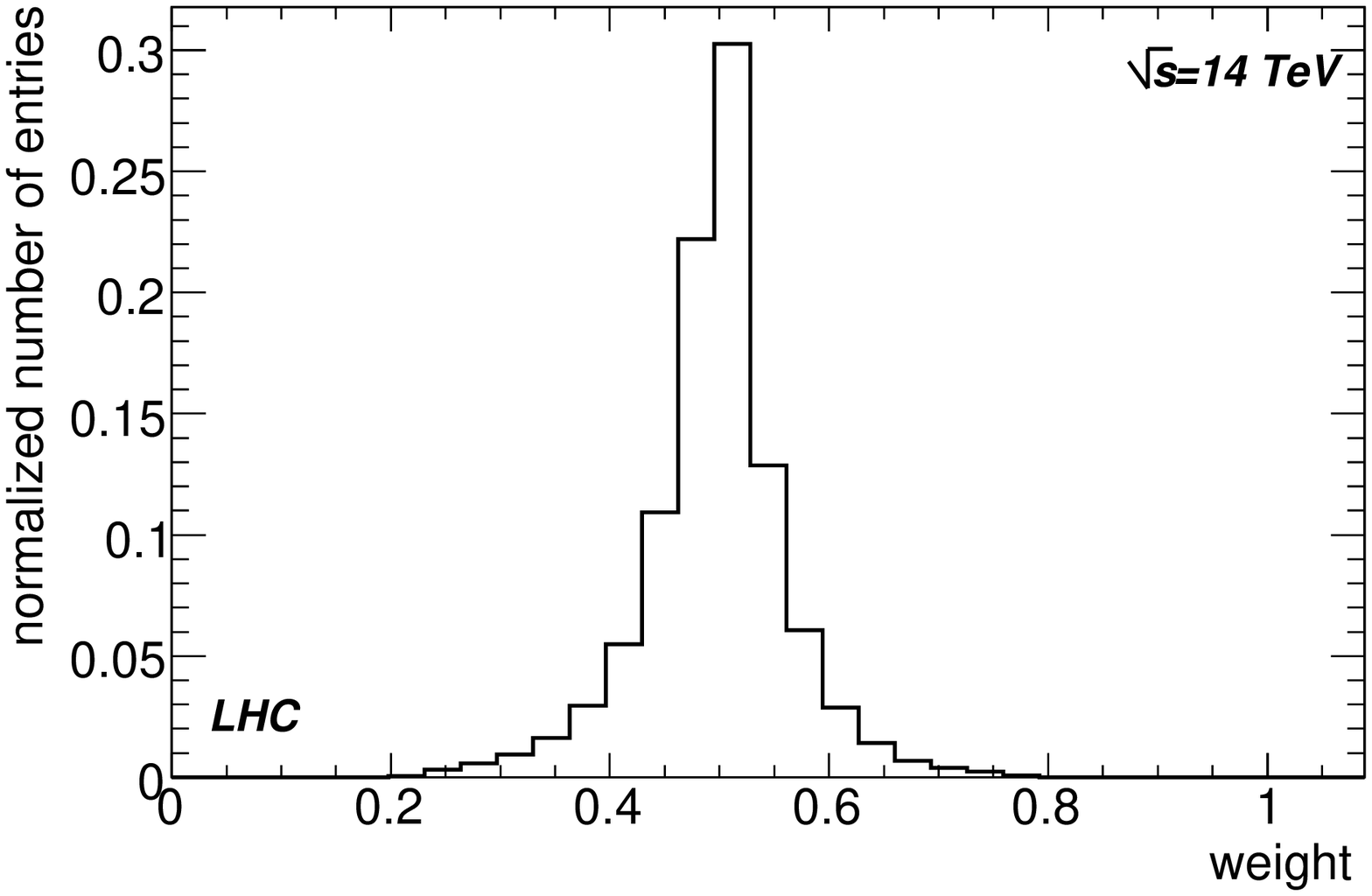}\\
\footnotesize {b) Weight given to the right solution for $W^+$}
\end{center}
\end{minipage}
\caption[Weight given to correct rapidity solution at the LHC]{The weight given to reconstructed rapidity solution closest to the true value of $y_W$ for $W^-$ and $W^+$ at the LHC.
}
\label{fig:lhc_rightweights}
\end{figure}

Figure \ref{fig:lhc_rightweights} displays the weight given to the rapidity 
solution that was closest to the correct rapidity at the LHC. While at the 
Tevatron, this distribution peaks at 1 and is mostly larger than 0.5 
(as shown in fig. \ref{fig:fermilab_rightweight}), at the LHC most weights are around
 0.5. For $W^-$ production the weights given to the right solution are on
 average at least marginally larger than 0.5, however  for $W^+$ rapidities it
 is narrow and symmetric around 0.5, thus reducing the impact of the weighting procedure. The larger spread of $W^-$ weights comes from the different expected $d\sigma/dy_W$ rather than decay angle effects. 

\begin{figure}
\begin{center}
\begin{minipage}[c]{0.45\textwidth}
\begin{center}
\includegraphics[width=0.99\textwidth]{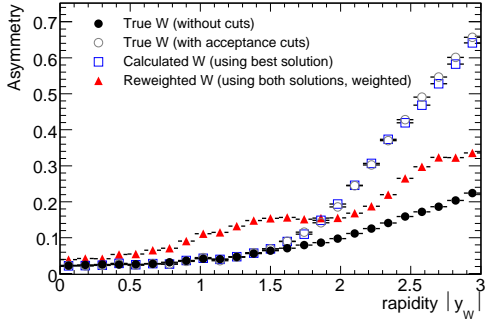}
\end{center}
\end{minipage} 
%%%%%%%%%%%%%%%%%%%%%%%%%%%%
\begin{minipage}[c]{0.45\textwidth}
\begin{center}
\includegraphics[width=0.99\textwidth]{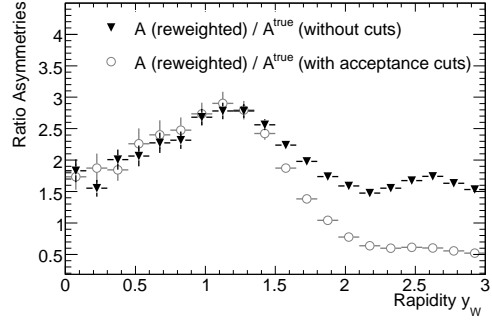}
\end{center}
\end{minipage} 
%%%%%%%%%%%%%%%%%%%%%%%%%%%%
\caption[Asymmetry at the LHC]{a)the asymmetry as evaluated at various stages of the iterative weighting procedure. b) the ratios of reconstructed asymmetry and true asymmetries (with and without cuts). 
\label{fig:lhc_acceptances_asy}}
\end{center}
\end{figure}

Again it is instructive to evaluate the asymmetry distribution at the various stages of the iterative weighting procedure. This is done in figure \ref{fig:lhc_acceptances_asy} a). Additionally, figure \ref{fig:lhc_acceptances_asy} b) depicts the ratio of the fully reconstructed asymmetry distribution to the true asymmetry distributions. The figure shows that in the central region, the corrections for intrinsic systematic biases of the full kinematic reconstruction dominates, while in the forward region genuine acceptance corrections dominate. Comparing this figure to the equivalent plots for the Tevatron (figure \ref{fig:fermilab_weighting2} b) reveals that the acceptance corrections are more important at the LHC than at the Tevatron. While at the Tevatron the ratio of the reconstructed asymmetry and the true asymmetry with cuts is between 0.8-1.1, the same ratio is 0.5-3.0 at the LHC. Equally the ratio of the reconstructed asymmetry and the true asymmetry without cuts is, at 1.5-3.0, larger at the LHC than at the the Tevatron (0.8-1.8).

\section{Modification of the analysis technique for the LHC}
\label{sec:NewScheme}

At the LHC the full $W$ kinematic reconstruction suffers because in this kinematic region it is difficult to favour one rapidity solution over another. Even with the best possible weights, it would not possible to restore the true asymmetry perfectly. At both the Tevatron and the LHC, acceptance corrections  compensate for intrinsic kinematic biases in the full kinematic reconstruction. As a result of these features a new simplified analysis technique for the LHC can be proposed: in this new simplified approach, the construction of weights for the full kinematic reconstruction is bypassed and instead  each solution is filled with a weight of 0.5, equivalent to randomly choosing either solution. This approach relies then solely on the acceptance correction, which 
is then also the only step dependent on MC input. This simplifies the determination of the systematic errors of the method. The acceptance corrections should be  applied in bins of $y_W$ and $p_T^W$. The iteration procedure with reweighting of the MC input is still necessary to remove the MC dependence. This approach is hereafter referred to as the ``new scheme''.

\section{Performance of the modified analysis technique at the LHC}
\label{sec:PerformanceNewScheme}

The expected systematic uncertainties of a measurement in the new scheme were evaluated in comparison to the uncertainty on the predicted $W$-asymmetry  and in comparison to the lepton asymmetry. Here, only the expected statistical errors and the inherent uncertainty of the modified $W$ measurement method are considered, other potential sources of experimental uncertainty are not taken into account. 

\subsection{Uncertainties on the acceptance corrections due to statistical fluctuations}

It is not trivial to propagate the uncertainties from statistical fluctations in the data sample and the acceptance corrections to a total statistical uncertainty on the $W$ asymmetry. The method reshuffles the number of events in the bins and might amplify statistical fluctuations as was also observed for the original implementation of the method~\cite{Han:2008zzc}. 

In order to estimate  the statistical uncertainties on the $W$ asymmetry, 450 toy experiments were carried out (375 in the case where the toy experiments contained $5 \times 10^6$ events each). Two {\sc Pythia} samples generated with the CTEQ66 PDF, $\sqrt{s}=14\,$TeV 
were used. The MC input sample consisted of 18$\times 10^6$ events and was used to calculate the acceptance corrections. 450 sub samples of $1\times10^5$, $5\times10^5$, $1\times 10^6$ and $5 \times 10^6$ events each were drawn randomly from the pseudo-data sample of 75$\times 10^6$ events, corresponding to integrated luminosities of 5.4, 27, 54 and 270 $\,\mathrm{pb}^{-1}$ respectively. These sub-samples were not exclusive, no attempt was made to exclude events already used in one toy data sample from other toy samples of the same luminosity.

\begin{figure}
%%%%%%%%%%%%%%%%%%%%%%%%%%%%
\begin{minipage}[c]{0.5\textwidth}
\begin{center}
\includegraphics[width=0.9\textwidth]{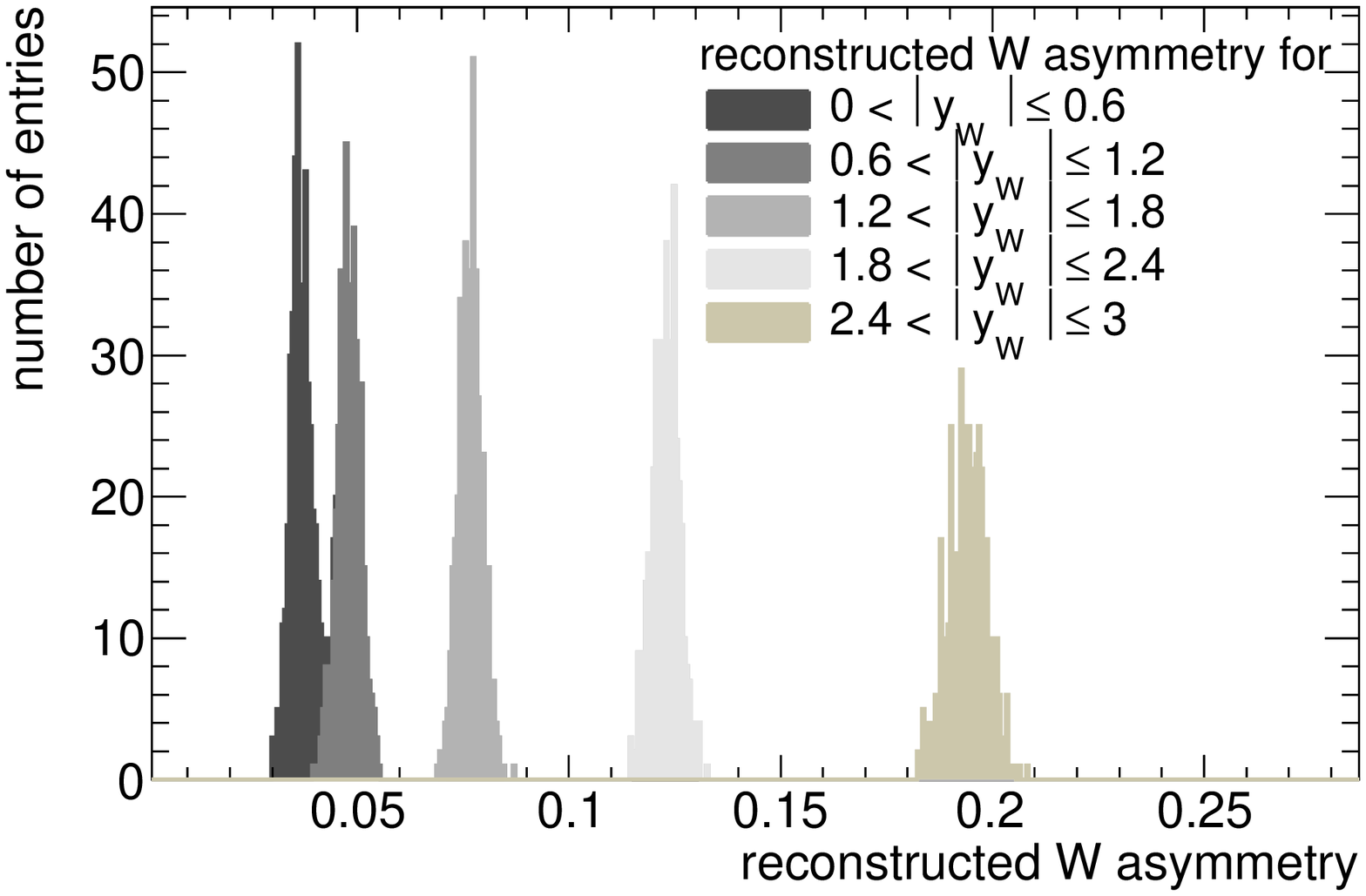}\\
\footnotesize {a) Number of entries for each measured W asymmetry (for a toy MC with 1~000~000 events).}\\
\end{center}
\end{minipage}
%%%%%%%%%%%%%%%%%%%%%%%%%%%%
\parbox{0.49\textwidth}{
\center
\begin{minipage}{0.4\textwidth}
\caption[Uncertainties on the acceptance corrections]{ a)  the number of entries for each measured W asymmetry. b) The statistical uncertainty corresponding to the spread as  an absolute uncertainty on the asymmetry. c) The statistical uncertainty as a relative uncertainty on the asymmetry.\\}
\label{fig:stat_error}
\end{minipage}}
\begin{minipage}[c]{0.5\textwidth}
\begin{center}
\includegraphics[width=0.9\textwidth]{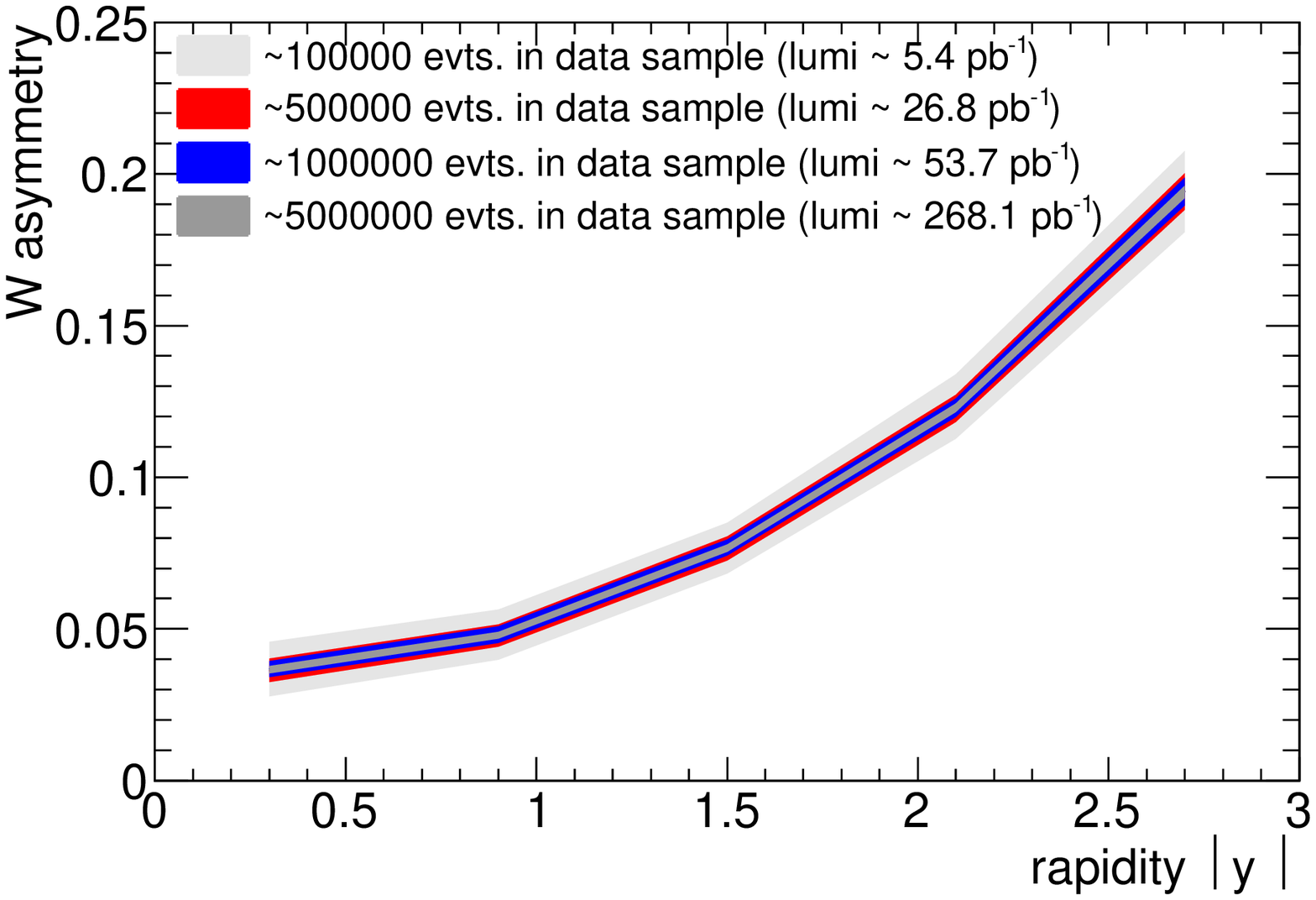}\\
\footnotesize {b) $W$ asymmetry with statistical uncertainty}
\end{center}
\end{minipage} 
%%%%%%%%%%%%%%%%%%%%%%%%%%%%
\begin{minipage}[c]{0.5\textwidth}
\begin{center}
\includegraphics[width=0.9\textwidth]{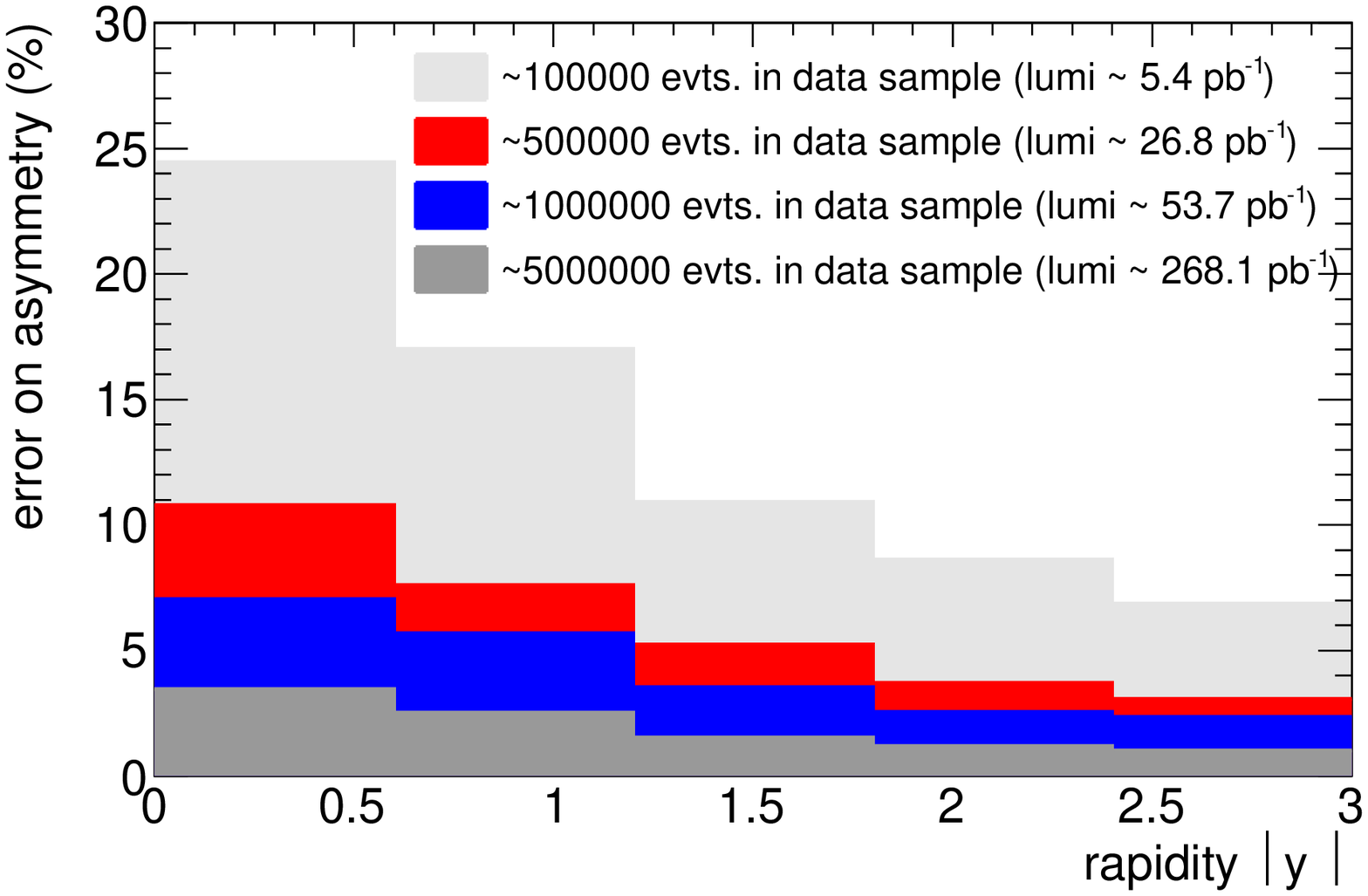}\\
\footnotesize {c) Relative uncertainty on $W$ asymmetry}
\end{center}
\end{minipage} 
\end{figure}

For each of the toy ``data'' samples, the $W$ rapidity solutions were calculated
 for each event and filled with a weight of 0.5. Next the acceptance 
corrections derived from the MC input sample were applied. The  spread of the 
 measured asymmetries in each bin of $y_W$ was then interpreted as the statistical uncertainty on the measured $W$ asymmetry. Figure \ref{fig:stat_error}
 a) shows this spread for the toy 'data' samples. 
 The entries belonging to different bins in $y_W$ are shown in different grey 
tones.
Figure \ref{fig:stat_error} b) depicts the $W$ asymmetry with the uncertainty
 extracted in the toy experiments as bands for the different integrated 
luminosities. Figure c) shows the  relative uncertainties on the $W$ asymmetry as a function of $y_W$.

\subsection{Uncertainites on the acceptance corrections from PDFs}

Uncertainties on the acceptance correction arising from the chosen   MC 
$y_{W^{\pm}}$ input distributions  differing from those in data due to the 
underlying PDF distributions were also considered. In this study the CTEQ66 
and MSTW08 PDF distributions were used. To estimate the uncertainty the
 PDF error sets based on the standard Hessian technique~\cite{Pumplin:2002vw}
 were used.

\begin{figure}
%%%%%%%%%%%%%%%%%%%%%%%%%%%%
\begin{center}
\includegraphics[width=0.49\textwidth]{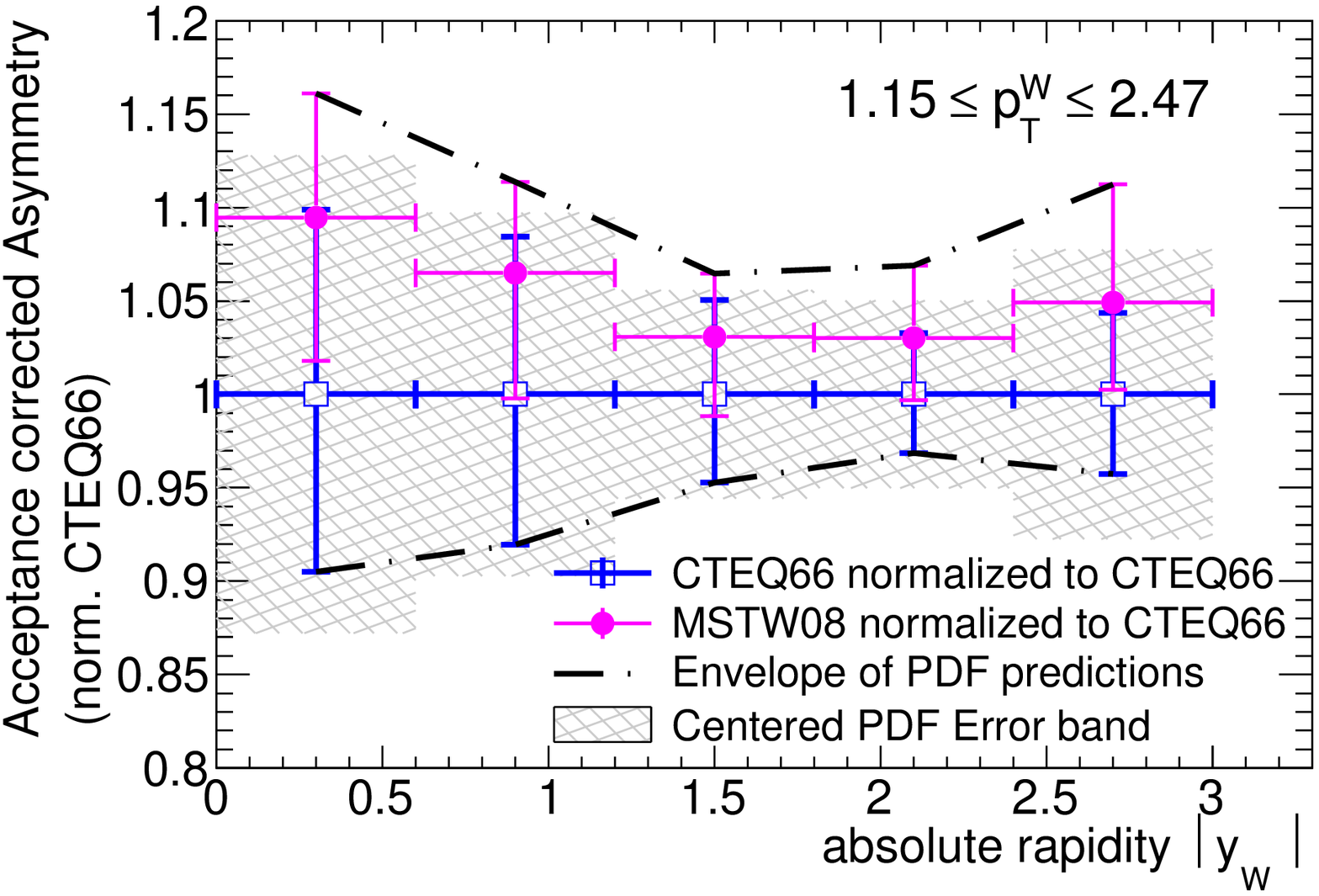}
\includegraphics[width=0.49\textwidth]{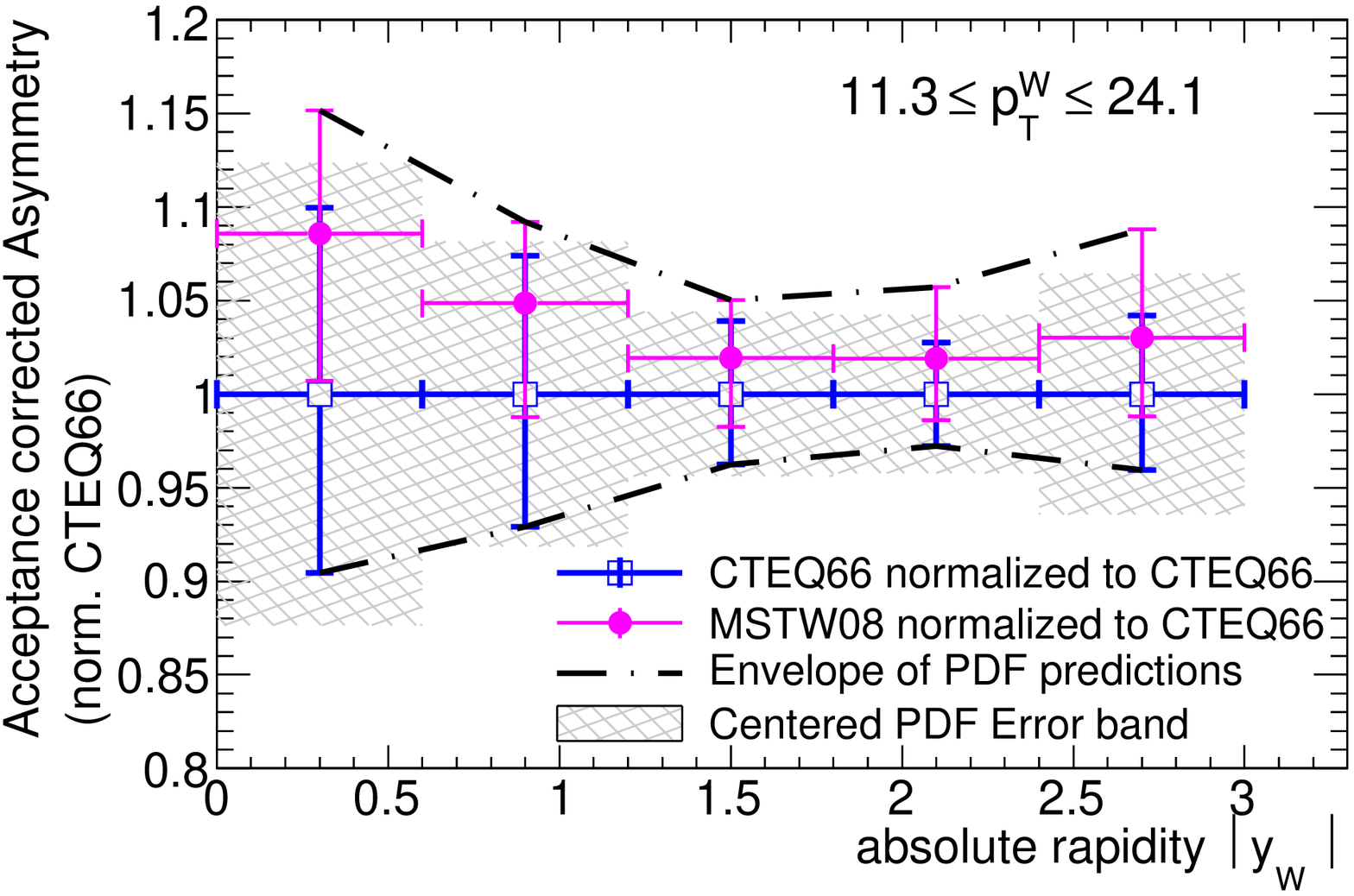}
\end{center}
\caption[Error on $W$ ]{The uncertainties on the measured W asymmetry from statistics and from the PDF uncertainties in two different bins of $p_T^W$. \label{fig:accPDFerrors}}
\end{figure}

 Figure \ref{fig:accPDFerrors} shows the measured asymmetries and the
 uncertainty on them arising from PDF uncertainties  for the CTEQ66 
 and the MSTW08 PDF sets, both shown as a ratio to the central value of 
CTEQ66. Around the PDF predictions and their uncertainties, a dot-dashed envelope is drawn,
 which is taken to give the PDF uncertainty on the  asymmetry measurement due
 to 
the acceptance  corrections. This error band is also centred around the 
CTEQ66 value and drawn as a grey band. Figure \ref{fig:accPDFerrors} shows
 on the left the relative PDF uncertainty on the asymmetry for 
$1.15<p_T^W<2.47\,$GeV and for $11.3<p_T^W<24.1\,$GeV on the right. For 
these and also the other low and mid-$p_T^W$ bins the uncertainties are 
between  5-12\%, for larger values of $p_T^W$, the uncertainty tends to 
decrease. 

\subsection{Experimental uncertainty due to the detector}

In order to estimate the effect of the reconstruction of the $W$ events in the detector, the four-vectors of the generator level particles were smeared to approximate the effect of differing MC and data resolutions. This was done by applying a Gaussian resolution factor to the $p_x$ and $p_y$ variables of the neutrino and to the $p_x$, $p_y$ and $p_z$ variables of the lepton. These resolution factors were drawn randomly and independently for each of the variables from a Gaussian distribution centered around one with a width of 1\% (lepton) and 5\% (neutrino) which represent
realistic uncertainties on the resolutions at the LHC~\cite{Lohwasser:2009zz}.
This was done 500 times for the same $\approx$ $16\times 10^6$ {\sc Pythia} events. The reconstructed asymmetry was corrected with the nominal acceptance corrections derived from the unsmeared sample. In a sizeable number of events, as a result of the smearing, no physical solution could be obtained for $y_W$. In these cases, the neutrino $p_T$ is scaled down in steps of 0.1 GeV until a valid solution is found, following the prescription of the CDF measurement, where the $\displaystyle{\not} E_T$ was assumed to be overestimated and scaled down~\cite{Aaltonen:2009ta}. 

Due to the fact that the $p_T$ spectrum of leptons is correlated with $\eta$, there is a slight bias in the mean of the smeared lepton asymmetry distributions with regard to the nominal (unsmeared) distribution. A similar but larger bias can also be observed in the measured $W$ asymmetry distribution. 
When calculating the experimental uncertainty, the bias of the mean of the 500 smeared distributions with regard to the nominal distribution is added in quadrature to the RMS of the 500 smeared distributions for both the lepton and the $W$ asymmetry. 

\subsection{Application of the modified scheme to pseudo-data}

In the following the weighting procedure of the modified scheme is investigated for the LHC environment. Two cases were studied:
\begin{enumerate}
\item \textbf{Different PDFs}: A {\sc Pythia} sample generated using the MSTW08 PDF ($8\times 10^6$ events) was used as pseudo-data and a {\sc Pythia} sample generated using the CTEQ66 PDF ($75\times 10^6$ events) was used as MC input. 5 iteration steps were used. 
\item \textbf{Higher orders}: An Herwig++ W production sample generated using the MSTW08 PDF ($1\times10^6$ events) was used as pseudo-data and a {\sc Pythia} sample generated using the MSTW08 PDF ($18\times10^6$ events) was used as MC input. 5 iteration steps were used. 
\end{enumerate}

\begin{figure}
%%%%%%%%%%%%%%%%%%%%%%%%%%%%
\begin{minipage}[c]{0.5\textwidth}
\begin{center}
\includegraphics[width=0.9\textwidth]{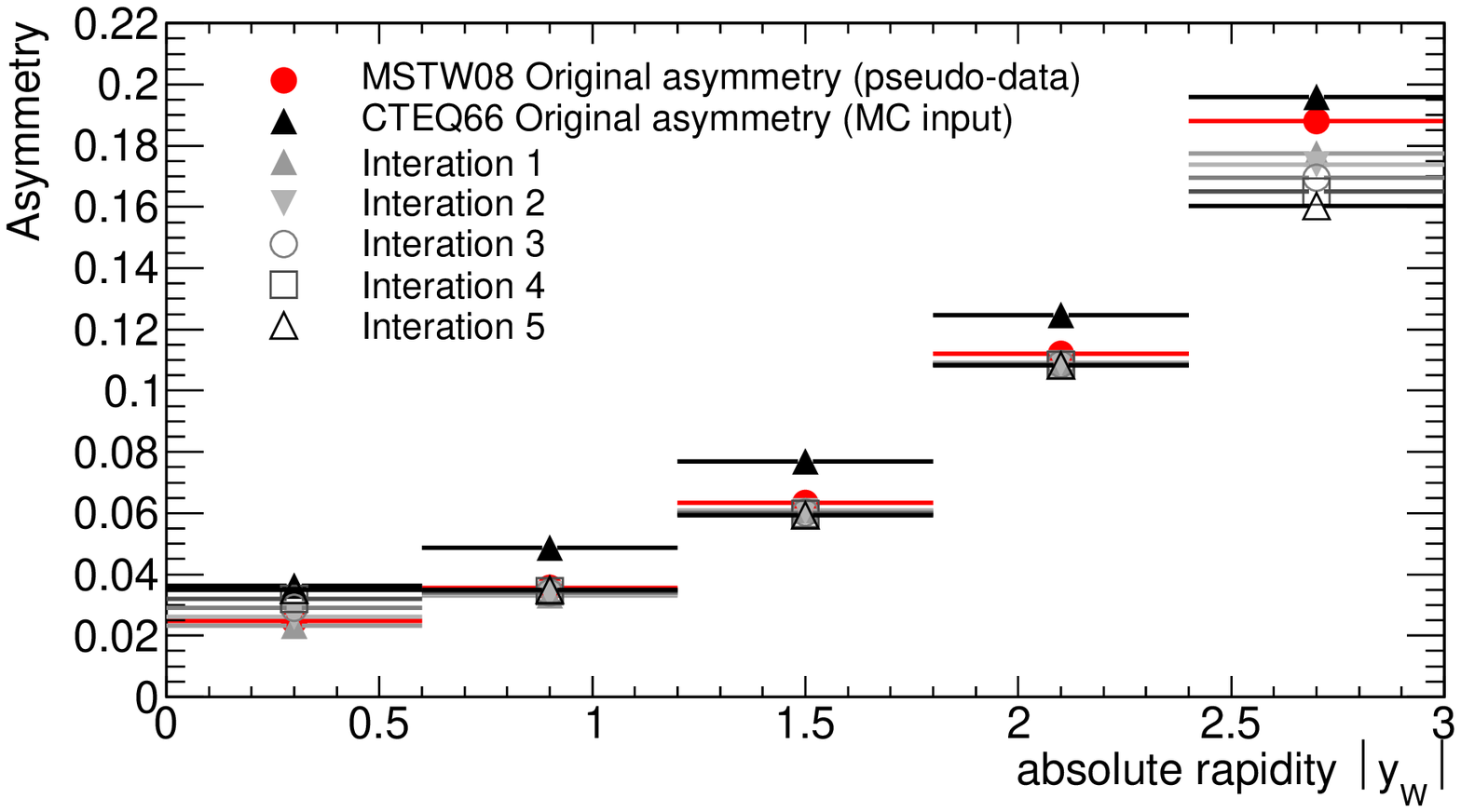}\\
\footnotesize {a) $W$ asymmetry of MC input, pseudo-data and interation steps}\\
\end{center}
\end{minipage}
%%%%%%%%%%%%%%%%%%%%%%%%%%%%
\parbox{0.49\textwidth}{
\center
\begin{minipage}{0.4\textwidth}
\caption[Performance of New $W$ reweighting scheme]{\footnotesize {The true $W$ asymmetries for LO MC input (CTEQ66 PDF, triangles) and pseudo-data (MSTW08, circles) are compared a) to the reconstructed $W$ asymmetries as obtained in the iterations steps 1-5. b) depicts the ratio of the reconstructed $W$ asymmetries of the iteration steps to the true $W$ asymmetry of the pseudo-data. c) shows the ratio of the reconstructed $W$ asymmetries of the iteration steps to the $W$ asymmetry of the preceeding iteration step. \\}}
\label{fig:application_pdfs}
\end{minipage}
}
\begin{minipage}[c]{0.5\textwidth}
\begin{center}
\includegraphics[width=0.9\textwidth]{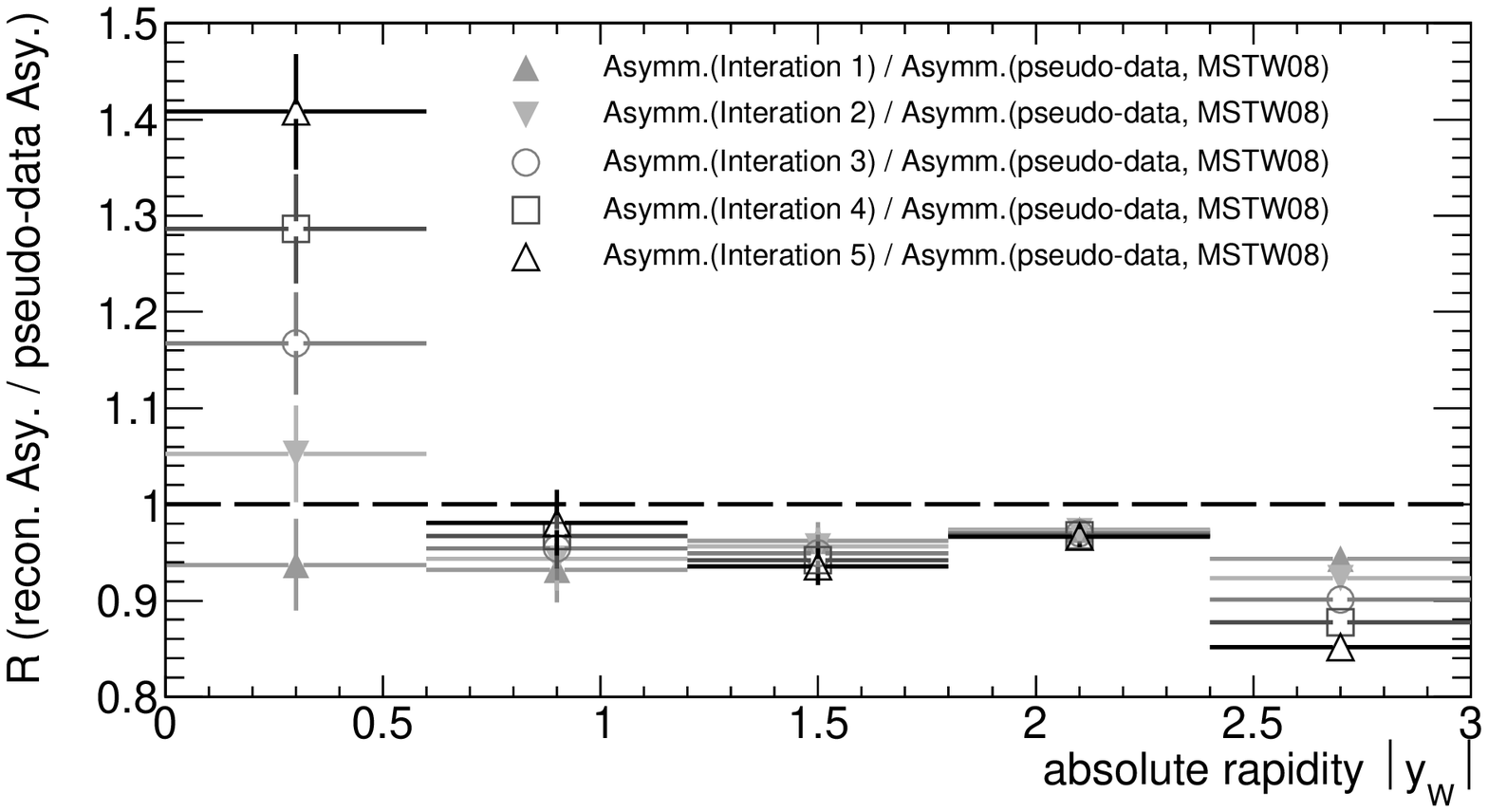}\\
\footnotesize {b)  Ratio of reconstructed $W$ asymmetry with the true $W$ asymmetry of the pseudo-data.   \\}
\end{center}
\end{minipage} 
%%%%%%%%%%%%%%%%%%%%%%%%%%%%
\begin{minipage}[c]{0.5\textwidth}
\begin{center}
\includegraphics[width=0.9\textwidth]{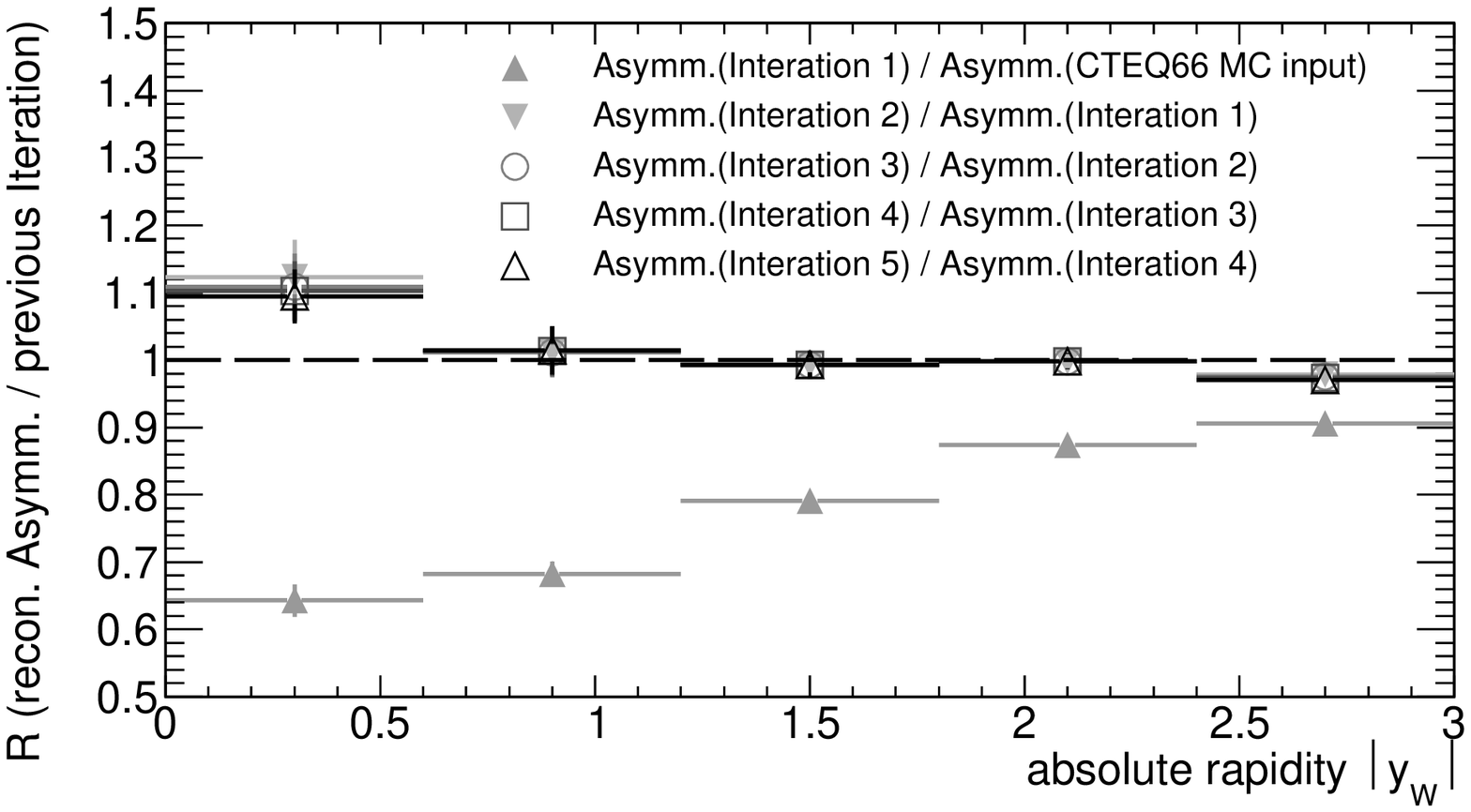}\\%asy_by_original_cteqmstw_highstat_new
\footnotesize {c) Ratio of reconstructed $W$ asymmetry with the reconstructed $W$ asymmetry of the preceding iteration step.\\}
\end{center}
\end{minipage} 
\end{figure}

Figure~\ref{fig:application_pdfs}~a) compares the true $W$ asymmetries for the
 CTEQ66 MC input and the MSTW08 pseudo-data to
 the reconstructed $W$ asymmetries as obtained in the iterations steps 1-5.
 The reconstructed $W$ asymmetries from all of the iteration steps agree
 better with the MSTW08 pseudo-data than with the CTEQ66 MC input, showing that the method still works. This can also be seen in figure~\ref{fig:application_pdfs}~b) which shows the ratio of the
 reconstructed $W$ asymmetries at each iteration step to the true 
$W$-asymmetry of the MSTW08 pseudo-data. The first iterations agree within 
7\%, for the other iterations larger deviations are visible. This is an effect
 of statistical fluctuations, the iteration procedure should be sensibly 
stopped after the first or second iteration for a measurement. This can be 
seen in figure~\ref{fig:application_pdfs}~c), which shows the ratio of the
 reconstructed $W$ asymmetries of the iteration steps to the $W$ asymmetry of
 the specific previous iteration step or the input MC asymmetry in the case of
 the first iteration). Except for the first iteration step, all of the other
 iteration steps agree very well with the reconstructed asymmetries from the 
preceeding iterations indicating the convergence of the iterative procedure.
 This demonstrates that the modified reweighting scheme is able to provide a 
good measurement of the $W$ asymmetry. 

An indication for when the iteration should be stopped can also be taken from 
this last plot: As long as the ratio of the reconstructed $W$ asymmetry of 
iteration step $n$ to the reconstructed $W$ asymmetry of iteration step $n-1$ 
grows significantly closer to one with every iteration $n$, the iteration 
still converges. If the ratio does not change significantly from some 
constant numbers (as e.g. in the first bin of $y_W$) or starts to deviate 
from one, the iteration starts to diverge and should be stopped well before 
this happens. 

%%%%%%%%%%%%%%%%%%%%%%%%%%%%

\begin{figure}
%%%%%%%%%%%%%%%%%%%%%%%%%%%%
\begin{minipage}[c]{0.5\textwidth}
\begin{center}
\includegraphics[width=0.9\textwidth]{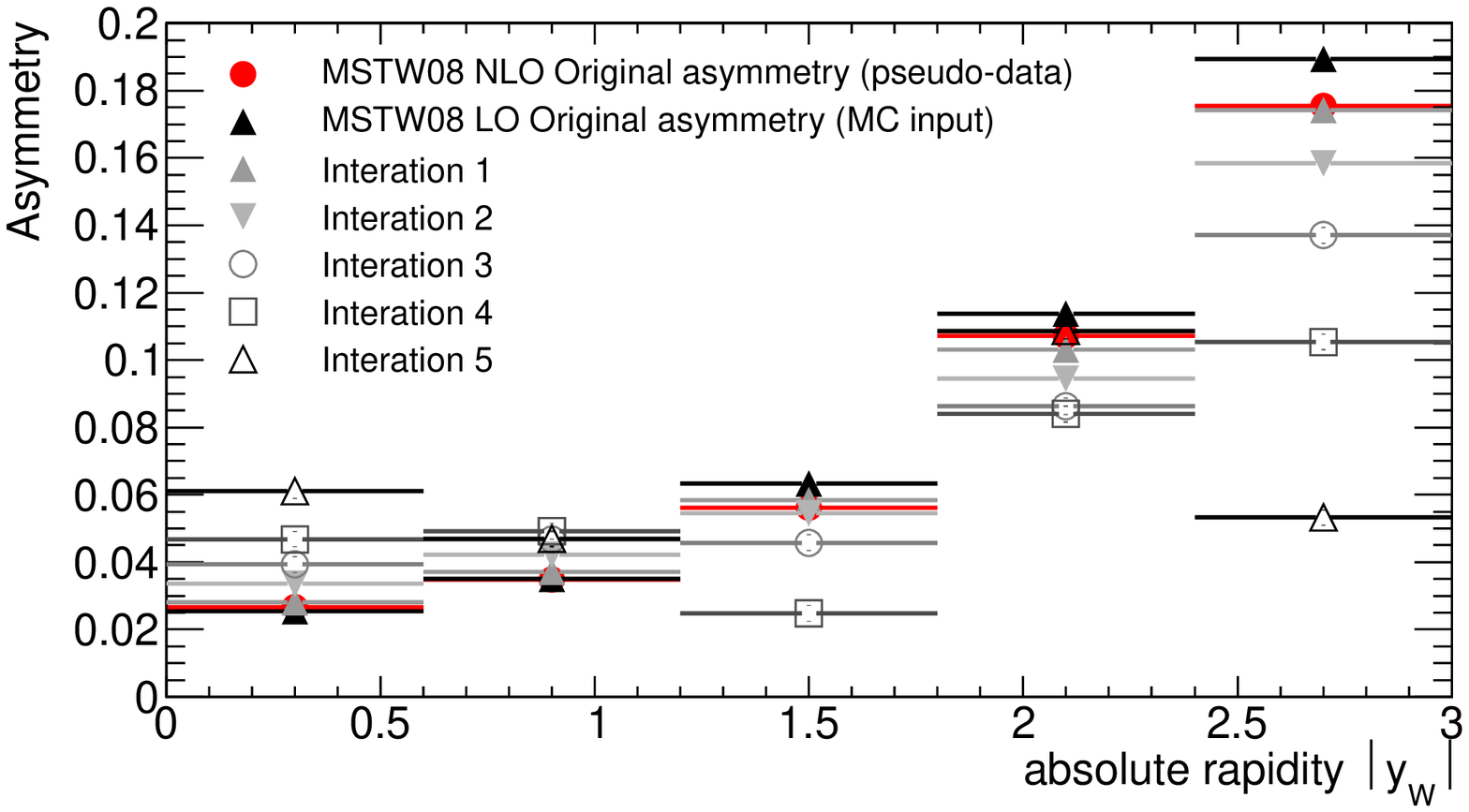}\\%Asy_Iter_powheg.eps}\\
\footnotesize {a) $W$ asymmetry of MC input, pseudo-data and interation steps}\\
\end{center}
\end{minipage}
%%%%%%%%%%%%%%%%%%%%%%%%%%%%
\parbox{0.49\textwidth}{
\center
\begin{minipage}{0.4\textwidth}
\caption[Performance of New $W$ reweighting scheme, higher order effects]{\footnotesize {The true $W$ asymmetries for MC input (LO, triangles) and pseudo-data (NLO, circles) are compared in a) to the reconstructed $W$ asymmetries as obtained in the iterations steps 1-5. b) depicts the ratio of the reconstructed $W$ asymmetries of the iteration steps to the true $W$ asymmetry of the pseudo-data. c) shows the ratio of the reconstructed $W$ asymmetries of the iteration steps to the $W$ asymmetry of the preceeding iteration step.\\}}
\label{fig:application_higherorder}
\end{minipage}
}
\begin{minipage}[c]{0.5\textwidth}
\begin{center}
\includegraphics[width=0.9\textwidth]{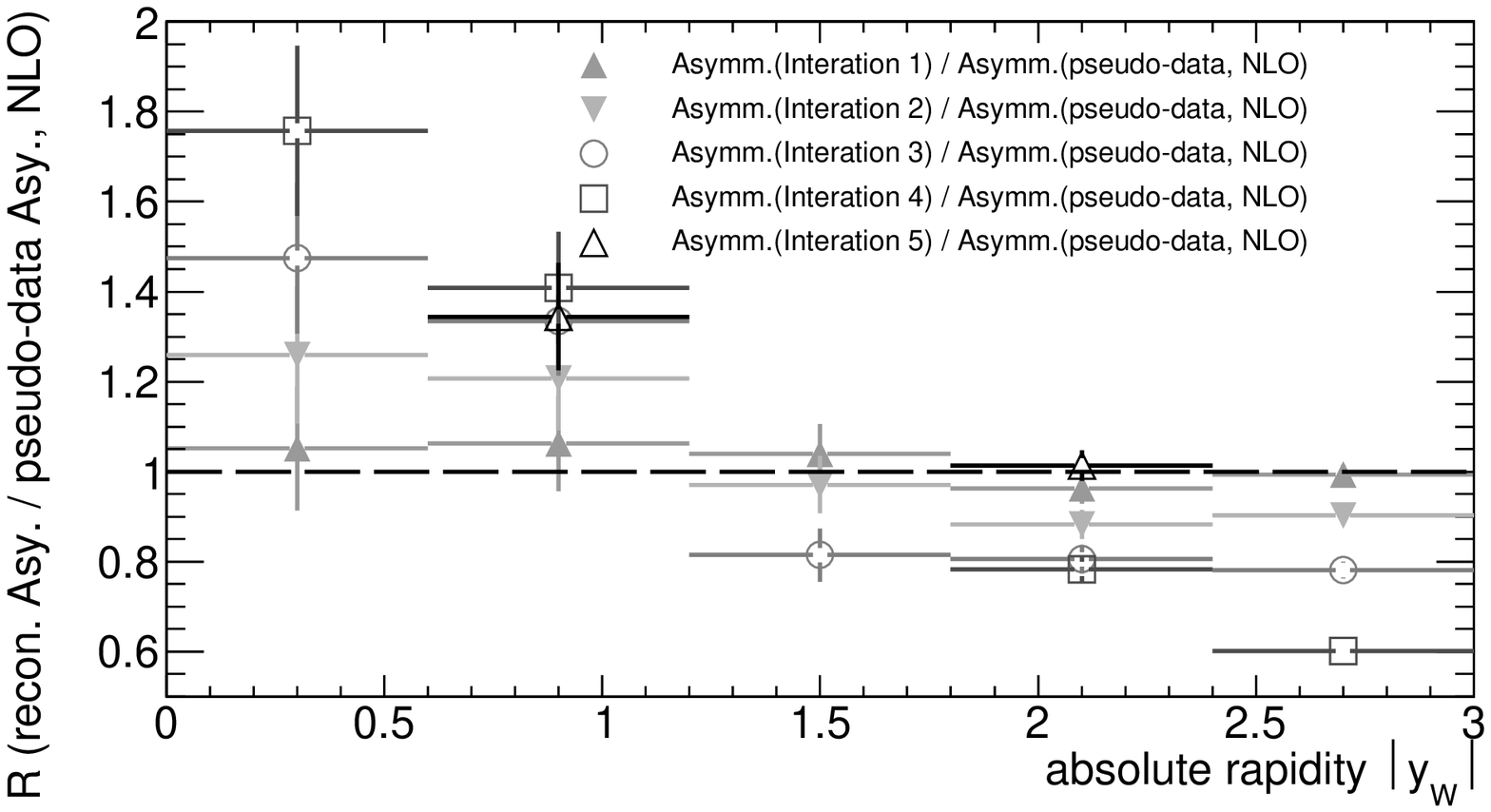}\\%AsyByOri_Iter_powheg.eps}\\
\footnotesize {b)  Ratio of reconstructed $W$ asymmetry with the true $W$ asymmetry of the pseudo-data.  \\ }
\end{center}
\end{minipage} 
%%%%%%%%%%%%%%%%%%%%%%%%%%%%
\begin{minipage}[c]{0.5\textwidth}
\begin{center}
\includegraphics[width=0.9\textwidth]{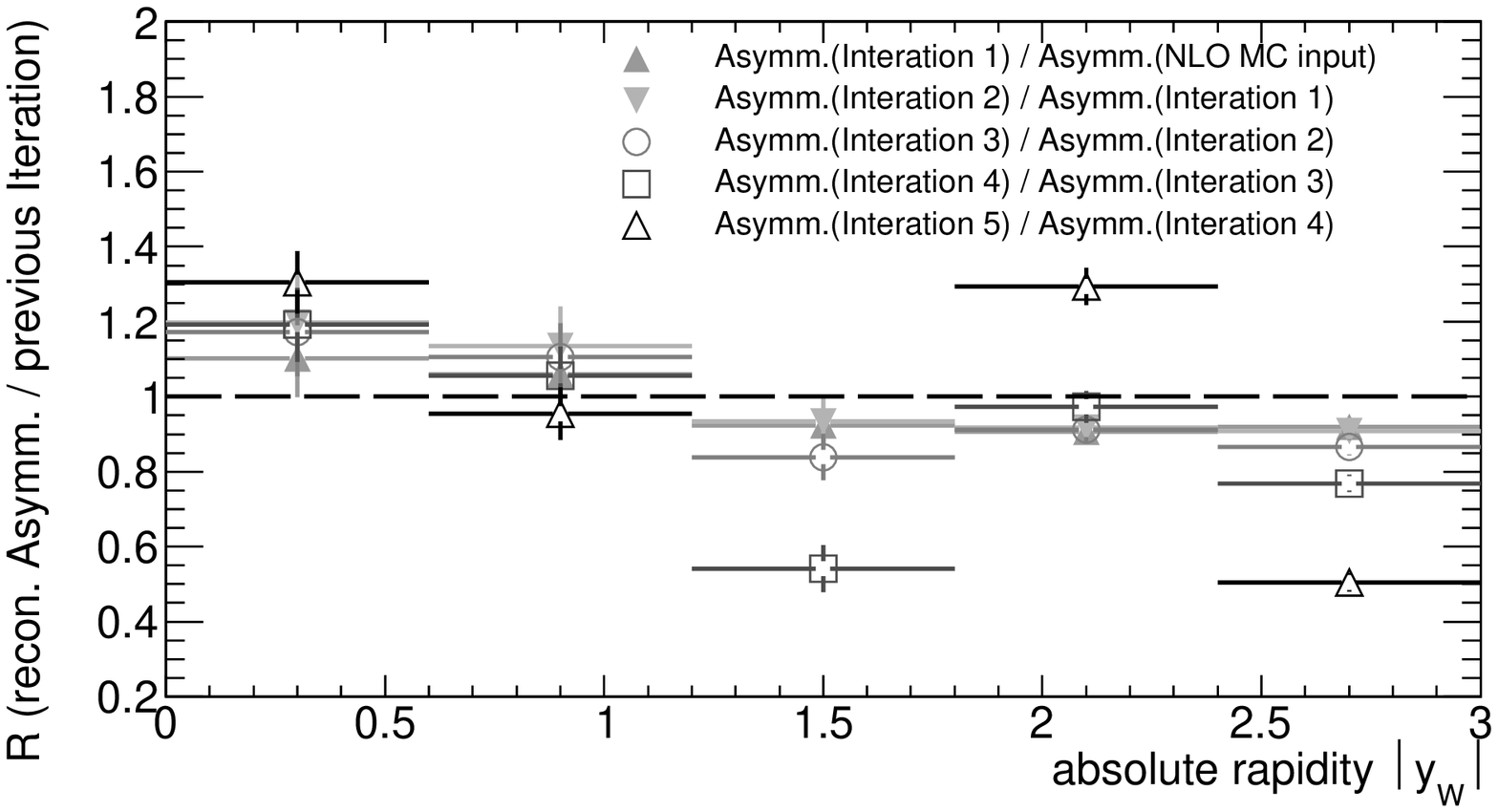}\\
\footnotesize {c) Ratio of reconstructed $W$ asymmetry with the reconstructed $W$ asymmetry of the preceding iteration step.\\}
\end{center}
\end{minipage} 
\end{figure}

%%%%%%%% NLO EFFECTS
All MC codes only simulate the physics process up to a certain order in QCD. The impact of higher order QCD effects was investigated by studying the analysis technique using NLO pseudo-data generated with Herwig++ and LO MC input generated with {\sc Pythia} each using the same PDF set.  Figure~\ref{fig:application_higherorder}~a) compares the true $W$ asymmetries  for the LO MSTW08 MC input and the NLO MSTW08 pseudo-data  to the reconstructed $W$ asymmetries. Again, the reconstructed $W$  asymmetries agree well with the true $W$ asymmetry of the  pseudo-data, as also seen in figure~\ref{fig:application_higherorder}~b). The  agreement is of the level of 5\% for the first iteration. Statistical effects are larger due  to the size of the data sets (1 million events for the NLO MSTW08 pseudo-data and 18 million events for the LO MSTW08 MC input). This can also be seen in  figure~\ref{fig:application_higherorder}~c), where the deviations of the iteration  step from the previous iteration step are quite small for the first and the  second iteration and then become systematically larger. Since this indicates a divergence, that grows with the number of iteration steps, the iteration should be stopped after the first or second iteration.   Apart from the large statistical fluctuations, no significant systematic effects were encountered from higher order QCD. 

%%%%%%%%%%%%%%%%%%%%%%%%%%%%

\section{Comparison of Uncertainties on the Lepton and $W$ asymmetry}\label{sec:comparisons}

In order to assess the possible advantage of the direct reconstruction of the $W$ asymmetry in comparison to a measurement of the lepton asymmetry, the expected statistical and systematic uncertainties on measurements of the two observables were compared to the uncertainty on the respective predictions arising from the PDFs. Here, for the $W$ asymmetry, the new scheme as described in section \ref{sec:NewScheme} was used and acceptance cuts were applied. For the lepton asymmetry, the same acceptance cuts were applied as would be the case for the usual experimental determination of this quantity.

Table \ref{tab:errorcomparison} compares the relative uncertainties in detail for the $W$ and lepton asymmetries including experimental uncertainties estimated using Gaussian resolution functions of 1\% for the lepton $P^l_T$ and 5\% for the $\displaystyle{\not} E_T$. 

Table \ref{tab:errorcomparison2} compares the uncertainties on the theoretical predictions from the PDFs ($\sigma_{\mathrm{PDF}}$) to the total uncertainties on the measurement (both with and without the experimental uncertainty due to the detector resolution). Here, the measured $W$ asymmetry refers to the $W$ asymmetry extracted in the last step of the analysis technique (c.f section \ref{sec:analysis}, step 7) with the modified scheme described in section \ref{sec:NewScheme}. Table \ref{tab:errorcomparison2} also gives the ratios of the errors on the measurement,$\sigma^{\mathrm{meas}}_{\mathrm{tot}}$, to $\sigma_{\mathrm{PDF}}$ for the W asymmetry, $\mathcal{R}(W)$, and lepton asymmetry, $\mathcal{R}(l)$. These ratios give an estimate of 
how well a measurement could be used to constrain the PDFs. Also shown is the ratio of these ratios, $\mathcal{R}(W/l)=\mathcal{R}(W)/\mathcal{R}(l)$. This is an estimate of how the observables perform in comparison to each other, with a ratio of 0.3 predicting the $W$ asymmetry measurement to be only a third as well suited to constrain PDFs as the lepton asymmetry for all $y_W$ when the uncertainties without the experimental uncertainty are taken into account. The detailed numbers for different $y_W$ bins are given in table \ref{tab:errorcomparison} and \ref{tab:errorcomparison2}, in which they are calculated for 268.1 pb$^{-1}$ integrated luminosity and are the relative uncertainties. 

The numbers are crude estimates since only the generator level is studied. However, the message from table \ref{tab:errorcomparison2} is clear: the measurement of the lepton asymmetry will be much more appropriate to constrain the PDFs, regardless of whether only the inherent theoretical uncertainties from the method are considered or whether experimental uncertainties are also taken into account. Measuring the $W$ asymmetry has a power to constrain PDFs which is only 33\% as good as a measurement of the lepton asymmetry, when only the inherent theoretical uncertainties on the method are taken into account (c.f. $\mathcal{R}(W/l)\sigma_{\mathrm{PDF}}/\sigma^{\mathrm{meas}}_{\mathrm{th}}$). Taking into account experimental resolution effects, the lepton asymmetry still performs twice as well compared to the $W$ asymmetry measurement (c.f. $\mathcal{R}(W/l)\sigma_{\mathrm{PDF}}/\sigma^{\mathrm{meas}}_{\mathrm{exp}}$). The ratio $\mathcal{R}_W/\mathcal{R}_{\mathrm{l}}$ tends to decrease with higher luminosity. 

It should be noted again that the situation at the Tevatron is more favourable for the measurement of the $W$ asymmetry. Since the lepton asymmetry at the Tevatron tends to be smaller than the $W$ asymmetry, at the Tevatron the lepton asymmetry is favoured if the systematic uncertainties are negligible or of the same absolute size on the $W$ and the lepton asymmetry.

\begin{table}
  \begin{center}
\footnotesize
    \begin{tabular}{|l|c|c|c|c|c||c|c|c|c|c|}
      \hline
$y_W$ or $\eta_l$           &$\sigma_{\mathrm{stat}} (W)$   &$\sigma_{\mathrm{acc}} (W)$    &$\sigma_{\mathrm{det}} (W)$    &$\sigma^{\mathrm{meas}}_{\mathrm{th}} (W)$      &$\sigma^{\mathrm{meas}}_{\mathrm{tot}} (W)$    &$\sigma_{\mathrm{stat}} (l)$       &$\sigma_{\mathrm{det}}(l)$  &$\sigma^{\mathrm{meas}}_{\mathrm{tot}}(l)$         \\
\hline
0 - 0.6 &       3.47&   10.1&   9.61&   10.7&   14.4&   1.49&   4.04&   4.3 \\ \hline 
0.6 - 1.2 &     2.52&   8.31&   6.51&   8.68&   10.9&   1.24&   2.64&   2.92 \\ \hline 
1.2 - 1.8 &     1.54&   4.9&    0.699&  5.13&   5.18&   0.921&  1.46&   1.73 \\ \hline 
1.8 - 2.4 &     1.19&   3.3&    1.77&   3.51&   3.93&   0.695&  0.925&  1.16 \\ \hline 
2.4 - 3 &       1.02&   4.38&   3.96&   4.5&    5.99&   -&   -&    - \\ \hline 

\hline
    \end{tabular}
  \end{center}
  \caption[Error comparison]{Comparison of the relative uncertainties on the measured $W$ asymmetry and the lepton asymmetry for a sample corresponding to an integrated luminosity of $268\,\mathrm{pb^{-1}}$. The statistical uncertainty in the bins is denoted $\sigma_{\mathrm{stat}}$, uncertainties on acceptance corrections from PDFs are denoted $\sigma_{\mathrm{acc}}$ (c.f \ref{sec:analysis}, step 3). Uncertainties from detector resolution are denoted as $\sigma_{\mathrm{det}}$. The total uncertainty neglecting $\sigma_{\mathrm{det}}$ is labeled $\sigma^{meas}_{\mathrm{th}}$ and the total uncertainty including resolution effects $\sigma^{meas}_{\mathrm{tot}}$. }\label{tab:errorcomparison}
\end{table}

\begin{table}
  \begin{center}
\footnotesize
    \begin{tabular}{|l|c|c|c|c|c|c|c|c|}
      \hline
$y_W$ or $\eta_l$            &$\sigma_{\mathrm{PDF}} (W)$   &$\mathcal{R}$ ($W$)         &$\mathcal{R} (W)$        &$\sigma_{\mathrm{PDF}}(l)$   &$\mathcal{R} (l)$     &$\mathcal{R}(l)$          &$\mathcal{R}(W/l)$      &$\mathcal{R} (W/l)$      \\
        &       &$\sigma_{\mathrm{PDF}}/\sigma^{\mathrm{meas}}_{\mathrm{th}}$    &$\sigma_{\mathrm{PDF}}/\sigma^{\mathrm{meas}}_{\mathrm{tot}}$   &        &$\sigma_{\mathrm{PDF}}/\sigma^{\mathrm{meas}}_{\mathrm{th}}$   &$\sigma_{\mathrm{PDF}}/\sigma^{\mathrm{meas}}_{\mathrm{tot}}$          &$\sigma_{\mathrm{PDF}}/\sigma^{\mathrm{meas}}_{\mathrm{th}}$     &$\sigma_{\mathrm{PDF}}/\sigma^{\mathrm{meas}}_{\mathrm{tot}}$          \\ \hline

0 - 0.6 &      24.5&   2.29&   1.7&    11.9&   8&      2.77&   0.287&  0.616\\ \hline
0.6 - 1.2 &     20.7&   2.39&   1.91&   11.2&   9.07&   3.85&   0.263&  0.496\\ \hline
1.2 - 1.8 &     15.1&   2.95&   2.92&   9.69&   10.5&   5.6&    0.281&  0.521\\ \hline
1.8 - 2.4 &     11.7&   3.32&   2.97&   7.75&   11.1&   6.7&    0.298&  0.443\\ \hline
2.4 - 3 &       8.39&   1.87&   1.4&    -&      -&   -&   -&   -\\ \hline

\hline
    \end{tabular}
  \end{center}
  \caption[Error comparison]{Comparison of the relative uncertainties on the measured $W$ and lepton asymmetries and the uncertainties on the prediction arising from PDFs for a sample corresponding to an integrated luminosity of $268\,\mathrm{pb^{-1}}$. The uncertainties on the theoretical predictions for the W and lepton asymmetry are denoted as $\sigma_{\mathrm{PDF}}(W)$ and  $\sigma_{\mathrm{PDF}}(l)$ respectively.  Ratios of uncertainties  for $W$ and lepton asymmetries are denoted as $R(W)$ or $R(l)$ respectively. Finally double ratios $\mathcal{R}(W)/\mathcal{R}(l)$ are denoted as $\mathcal{R}(W/l)$.}\label{tab:errorcomparison2}
\end{table}

\section{Conclusions}\label{sec:conclusions}

In this paper the applicability of the analysis technique first developed for the Tevatron by Bodek {\it et al.}~\cite{Bodek:2007cz} to the LHC environment has been studied.  The method is inherently less well suited for LHC conditions and a modification to the method has been proposed. This modification was found to be more robust and relies on MC input only for one step. The statistical and systematic uncertainties on the measurement of the $W$ asymmetry were determined and compared to the PDF uncertainties on the prediction from theory. The expected experimental uncertainty on the measured $W$ asymmetry is smaller than the theoretical uncertainty. However, the expected experimental uncertainty on the lepton asymmetry is even smaller in comparison to its theoretical uncertainties and thus better suited to constrain PDFs at the LHC. An estimate of the likely impact of experimental uncertainties due to the detector resolution were taken into account and the lepton asymmetry was still found to perform better than the $W$ asymmetry by a factor of roughly 2. It is hence the opinion of the authors that early measurement at the LHC should focus on the lepton asymmetry rather than the direct $W$ asymmetry to constrain the PDFs. The studies have been performed for a centre of mass energy of $\sqrt{s}=$ 14 TeV. However, the main arguments hold also at $\sqrt{s}=$ 7 TeV, so that these conclusions are not likely to change with the centre of mass energy. 

\section*{Acknowledgements}

This work was supported by the UK STFC. The authors would like to thank Yeon Chung, Bo-Young Han, Kevin McFarland and Arie Bodek for helpful clarifications of implementation of the method at the Tevatron. It is a pleasure to thank Claire Gwenlan for useful comments on the manuscript and to thank Eva Halkiadakis and Junjie Zhu for information about the comparison of the Tevatron $W$-asymmetry to the lepton asymmetry. 

K.~Lohwasser gratefully acknowledges support from Deutscher Akademischer Austauschdienst (DAAD). \providecommand{\href}[2]{#2}\begingroup\raggedright\endgroup

\end{document}